\documentclass{iopart}

\usepackage{iopams}

\usepackage{epsf,graphicx}

\newcommand{\eff}{{\rm eff}}
\newcommand{\inter}{{\rm int}}
\newcommand{\tot}{{\rm tot}}

\newcommand{\la}{\langle}
\newcommand{\ra}{\rangle}
\newcommand{\lla}{\left\langle}
\newcommand{\rra}{\right\rangle}

\begin{document}

\title[Correlation functions of scattering matrix elements]
{Correlation functions of scattering matrix elements in microwave cavities 
 with strong absorption}

\author{R. Sch\"afer$^1$, T. Gorin$^2$, 
        T. H. Seligman$^3$, and H.-J. St\"ockmann$^1$}

\address{$^1$ Fachbereich Physik, Philipps-Universit\"at Marburg, Renthof 5,
D-35032 Marburg, Germany}

\address{$^2$ Theoretische Quantendynamik, Fakult\" at f\" ur Physik,
               Universit\" at Freiburg, Hermann-Herder-Str. 3, 
               D-79104 Freiburg, Germany}

\address{$^3$ Centro de Ciencias F\'{\i}sicas, Universidad Nacional 
Aut\'{o}noma de M\'{e}xico, Campus Morelos, C.~P. 62251, 
Cuernavaca, Morelos, M\'{e}xico}

\begin{abstract}

The scattering matrix was measured for microwave cavities with two 
antennas. It was analyzed in the regime of overlapping resonances. The
theoretical description in terms of a statistical scattering matrix 
and the rescaled Breit-Wigner approximation has been applied to this regime.
The experimental results for the auto-correlation function show that the
absorption in the cavity walls yields an exponential decay. This
behavior can only be modeled using a large number of weakly coupled
channels.
In comparison to the auto-correlation functions, the cross-correlation 
functions of the diagonal S-matrix elements display a more pronounced difference
between regular and chaotic systems.

\end{abstract}



\ead{stoeckmann@physik.uni-marburg.de}

\maketitle

\section{Introduction} \label{sec:intro}

The correlation hole as seen in the Fourier transform of the stick spectrum
of a closed Hamiltonian quantum system is particularly sensitive to long
range spectral correlations, typical for random matrix models.
This behavior in turn is indicative of chaos in the
classical equivalent of the system via the quantum chaos conjecture
\cite{Cas80,Boh84b}. 
Extensive work on the detection of such correlations using the correlation
hole has been done in molecular physics \cite{Lev86,Lom91,Lom93}, nuclear physics \cite{Lom94}, and in the 
analysis of spectra from microwave cavities \cite{Alt97b} and optical resonators \cite{Din02}. 
As the energy spectrum often is not available, Jost and Lombardi have early 
focused attention on the analysis of intensity spectra (see e.g. \cite{Jos86}).

The Fourier transform $\hat\sigma_\alpha(t)$ of the intensity spectrum
$\sigma_\alpha(E) = \sum_i \alpha_i^2 \; \delta(E-E_i)$ for a system with
eigenenergies $E_i$ and a state with amplitudes $\alpha_i$ yields
\begin{equation}
\hat c_A(t) = \left|\hat\sigma_\alpha(t)\right|^2 = \sum_{i} \alpha_i^4 +
\sum_{i\ne j} \alpha_i^2\;\alpha_j^2 \;\rme^{2\pi\rmi\, (E_i-E_j)\, t} \; .
\end{equation}
The inverse participation ratio $\sum_i \alpha_i^4$ is $3/(N+2)$ for a 
random state.
Averaging over a Gaussian 
orthogonal ensemble (GOE) in the large $N$-limit leads to 
$N\, \hat c_A(t)= 3-b(t)$. Here $N$ is the dimension of 
the matrices and $b(t)$ is the two-point form factor of the GOE \cite{Meh91}.
For the corresponding stick spectrum $1-b(t)$ is obtained instead.
For intensity spectra the correlation hole is thus reduced to $1/3$ of
its full value. This reduction makes the detection of the correlation hole
difficult.
Reference~\cite{Alt97b} discusses different ways to overcome this basic problem,
which becomes more acute if cross sections are considered.
Indeed, in the case of cross sections, the use of auto-correlation functions
was shown to be of very limited efficiency.

The cross correlations of independent intensity spectra, by contrast, display the full
correlation hole. Indeed, performing the GOE average we find, instead of an inverse participation ratio, the product of 
two vector norms, i.e.
\begin{equation}
\hat c_C(t) 
   = \hat\sigma_1^*(t)\, \hat\sigma_2(t) = \sum_i \alpha_i^2\; \beta_i^2 +
     \sum_{i\ne j} \alpha_i^2\;\beta_j^2 \;\rme^{2\pi\rmi\, (E_i-E_j)\, t} \; ,
\end{equation}
where $\beta_i$ refers to the component of the second state. Therefore the cross
 correlation behaves
as $1-b(t)$.

This simple fact has led to a detailed study of the possibility
to observe the correlation hole in correlation functions of total
and partial cross sections \cite{Gor02}. To allow a comparison with regular systems, the Poisson orthogonal ensemble (POE) was used. 
It combines random, statistically independent eigenvalues \cite{Ber77c} with orthogonally invariant eigenvectors \cite{Dit91}. The POE does not have the universal implications of the GOE, both because the assumptions about spectral statistics are less well founded, and because we easily may encounter situations of preferred coordinate systems. Nevertheless, it is the best random matrix model for integrability that is available.
Indeed cross correlations
prove to be the tool of choice to detect the
correlation hole.  Clearly the main interest of such an analysis
results when the total absorption, i.e. the sum over all transmission
coefficients, is fairly large, which implies that the average total width
$\Gamma$ is large compared to the mean level spacing $d$. 

In the present paper we analyze the total cross sections of several normal-conducting microwave resonators with two antennas, obtained from measurements of the scattering matrix (S-matrix) via the optical theorem. The spectra of the studied systems exhibit different types of statistics, ranging from POE to GOE behavior.
The wall absorption is significant and either comparable to or 
much larger than the transmission of the antennas. This leads us from
resonances with small overlap to such with very strong overlap. Absorption channels are not directly accessible to experiments.

Our experiments address two interesting and quite general questions:
On one hand, we test the use of cross-correlation functions
to identify the effect of correlations in the spectrum of a chaotic Hamiltonian
in the case of overlapping resonances. 
The results of reference \cite{Gor02} are compared with the data. 
On the other hand, we investigate whether
absorption has to be included in terms of many weak or few strong channels,
or whether the two cases cannot be distinguished.
For this purpose we extend the
results of \cite{Gor02}
to include an infinite number of weak channels, which we shall show to
cause an exponential decay of the correlation functions.

In section 2 we recall some basics of random matrix scattering
theory and some results of \cite{Gor02} that are essential to our analysis,
and we discuss the effect of a large number of channels with small
absorption. In the following section the experimental setup and the
studied billiards are explained. 
In sections 4 and 5 we shall see that the signatures of chaos are more pronounced in the cross correlation than in the auto correlation.
Further we show that a description of absorption
in terms of many channels is essential to obtain agreement with
the experiment even in cases where the total absorption is
of order one.

\section{\label{B} Basics of scattering theory and theoretical developments
                   for absorption channels}

There is an exact correspondence between the stationary classical
wave equation of an ideal quasi-two-dimensional microwave cavity and the
stationary Schr\" odinger equation for  a two-dimensional quantum billiard of the
same shape. This correspondence includes the scattering situation,
by taking the antennas explicitly into account \cite{Stoe99}, and ultimately
even absorption phenomena.

The S-matrix for this situation is frequently used
to describe resonant scattering arising e.g. in nuclear, atomic
or mesoscopic systems \cite{Guh98}:
\begin{equation}
S(E) = 1 - \rmi V^T \; \frac{1}{E-H_\eff} \; V \; , \quad 
H_\eff = H_\inter - \frac{\rmi}{2} \; V\; V^T \; .
\label{defS}\end{equation}
Here, $H_\inter$ is the Hamiltonian describing the closed billiard, and $V$
is a $N\times M$ matrix, which couples the $N$ interior wave functions
to $M$ decay channels. For each antenna we need one channel
or column vector, where the components $V_{ja}$ are proportional to the amplitude of
the billiard eigenfunction at the position of the antenna,
\begin{equation}
  V_{ja} \propto \Psi_j(\vec r_a) \; ,
\label{defVja}\end{equation}
provided that the diameter of the antennas is small compared to the wavelength.
This approximation may only be used
in regions far from any thresholds, and the proportionality
``constant'' typically varies slowly with frequency \cite{Stoe02c}.
If symmetry-equivalent positions of the antennas are avoided, then typically the column vectors of $V$ are approximately orthogonal to each other. Though not essential, this assumption simplifies the theoretical analysis.

It is convenient to work in the eigenbasis of the closed system $H_{\rm int}$.
Then the scattering matrix depends on the eigenvalues of $H_{\rm int}$ and on 
the coupling amplitudes, defined in equation~(\ref{defVja}). The analysis of the 
experiment is carried out in the framework of random matrix theory, where 
chaos is represented by a GOE and integrability by a POE. The two ensembles
differ only in the distributions of the eigenvalues, while orthogonal 
invariance implies in both cases that the columns of the matrix $V$ are
distributed according to the invariant measure of the orthogonal group. In practice we use independent 
random Gaussian variables for the matrix elements, an approximation which
becomes valid for large $N$.

We first present some elementary results of scattering theory using the notation
 of \cite{Gor02}. From equation~(\ref{defS}) the cross sections are derived as
\begin{equation}
\sigma_{ab}(E) = \left|\delta_{ab} - S_{ab}(E)\right|^2 \; .
\label{defsigma}\end{equation}
Note that the experimental setup allows to measure $S_{ab}$ directly.
This avoids the difficulties related to the measurement of
total cross sections. The optical theorem establishes a linear relation between
 the
S-matrix elements and the total cross sections:
\begin{equation}
\sigma_\tot^{(a)}(E) = 2\; (1-{\rm Re}\; S_{aa}) \; .
\label{optthm}\end{equation}

The first quantity to study is the average S-matrix. The average
can be a spectral average (denoted by $\la\ldots\ra$), an ensemble average
$(\,\overline{\rule{0mm}{1.5ex}\ldots}\, )$, or a combination of both. For the 
average S-matrix all these averages must coincide. In \cite{Mel85} it has been
 shown that
\begin{equation}
\overline{S_{aa}(E)} = \frac{1-\kappa_a}{1+\kappa_a} \; ,
\label{eq:kappa}\end{equation}
if $\overline{S(E)}$ is real and diagonal in the block of observable
channels.
The $\kappa_a$ are real and positive parameters and relate to the coupling
 constants $V_{ja}$ and the
transmission coefficients $T_a$, respectively, via
\begin{equation}
\kappa_a = \frac{\pi}{2\, d} \la V_{ja}^2\ra \; , \quad
T_a = \frac{4\, \kappa_a}{(1+\kappa_a)^2} \; .
\label{defT}\end{equation}

The analysis will be performed in terms of correlation functions in the time
 domain,
\begin{equation}
\fl
\hat C[\sigma_\tot^{(a)},\sigma_\tot^{(b)}](t) = \frac{1}{L}\left\{
\lla \hat\sigma_\tot^{(a)}(-t)\; \hat\sigma_\tot^{(b)}(t) \rra -
\lla\hat\sigma_\tot^{(a)}(-t)\rra\;\lla\hat\sigma_\tot^{(b)}(t)\rra
\right\} \; ,
\label{defCtot}\end{equation}
where $\hat \sigma$ are Fourier transforms of the cross sections.
The Fourier transforms are taken over a window of size $L$. It should
be large compared to the average level distance $d$ but sufficiently small so that the average S-matrix may be assumed constant.

Note that due to the optical theorem (\ref{optthm}) the correlation function
between total cross sections is equal to the correlation function between
the corresponding S-matrix elements:
\begin{equation}
\hat C[\sigma_\tot^{(a)},\sigma_\tot^{(b)}](t)
   = \hat C[S_{aa},S_{bb}^*](|t|) \; .
\label{Csigtot}\end{equation}
For the GOE the correlation function can be calculated exactly, as will be shown
 below. For the general case we rely on the rescaled Breit-Wigner approximation (RBWA)
\cite{Gor02}.

The rescaling is necessary, as soon as the transmission
coefficients are not extremely small. In the standard Breit-Wigner
approximation, the average width of the resonances is given by:
\begin{equation}
\la\Gamma\ra = \frac{2\, d}{\pi} \sum_{c=1}^M \kappa_c \; .
\end{equation}
Yet Ericson showed that in the limit of
many channels of comparable coupling strength, the correlation function of
S-matrix elements or cross sections is proportional to $\exp(-\Gamma_C\, t)$.
According to this derivation \cite{Eri66,Bro81} $\Gamma_C$ should be equal
to $\la\Gamma\ra$, but actually
\begin{equation}
\Gamma_C = \frac{d}{2\pi} \sum_{c=1}^M T_c \;
\end{equation}
gives the correct value for the correlation width. The two expressions
coincide only in the limit where $\sum_{c=1}^M T_c\ll 1$.
Thus the standard Breit-Wigner approximation is not valid in the range we
are interested in. Fortunately, it turns out that the rescaling
$\kappa_c\to T_c/4$ compensates for this defect up to rather strong overlaps
\cite{Gor02}.
The rescaled Breit-Wigner approximation has already been applied in the regime of non-overlapping resonances \cite{alh98}, but without mentioning the difference to the usual Breit-Wigner result.

In the case of the correlation functions~(\ref{Csigtot}), we
first write the S-matrix elements~(\ref{defS}) in terms of the standard
Breit-Wigner approximation. Then we plug this into equation~(\ref{defCtot}) and
average over the spectrum of $H_0$. At last we perform the rescaling to obtain
\begin{equation}
\fl 
\hat C[\sigma_\tot^{(a)},\sigma_\tot^{(b)}](t) = T_a T_b \left\{
\lla g_a g_b \; \rme^{-G\; t}\rra
- \lla g_a\; \rme^{-G\; t/2} \rra \; \lla g_b\; \rme^{-G\; t/2} \rra \;
b_2(t) \right\} \; ,
\label{CstRBW}\end{equation}
with $G = \sum_{c=1}^M T_c \; g_c$. The normalized squared amplitudes
$V_{ia}^2/\la V_{ia}^2\ra$ are replaced by random variables $g_c$. Due to
the orthogonal invariance of the ensembles considered, these variables
are assumed to be Porter-Thomas distributed \cite{Bro81}. Note that
different situations may occur in the case of symmetries or integrable
dynamics.

We account for wall absorption by introducing $M_W$ additional channels. The
transmission summed over all of them must equal the wall absorption,
\begin{equation}
T_W= \sum_{c=1}^{M_W} T_{M_A+c} \; ,
\end{equation}
where $M_A=2$ is the number of antennas. 
If $M_W$ is small, the correlation functions depend on all transmission 
coefficients individually, and no simplification is possible. However, if $M_W$ 
is large, we may use that asymptotically
\begin{equation}
\fl 
\hat C[S_{ab},S_{cd}^*](t) \to \rme^{-T_W\, t}\; \hat C[\tilde S_{ab},\tilde S_{cd}^*](t)
   \quad\mbox{as}\quad  M_W\to\infty \; , \quad \max_{c>M_A}(T_c)\to 0 \; ,
\label{Wabsorp}\end{equation}
where $\tilde S_{ab}$ describes the scattering system with only $M_A$ channels, obtained
by eliminating the last $M_W$ columns in the coupling matrix $V$ (see 
equation~(\ref{defS})).
For the rescaled Breit-Wigner result~(\ref{CstRBW}) this proposition
follows from the central limit theorem.

For later use we produce the result for the correlation function in the case of two
antennas with equal transmission coefficients $T_1=T_2=T_A$. If the absorption in the
walls is taken into account in the many channel limit via equation~(\ref{Wabsorp}), 
the ensemble averages in equation~(\ref{CstRBW}) involve two random variables $g_1$ and 
$g_2$ only. Assuming that both are uncorrelated Porter-Thomas variables, and using that 
$G= T_A\, (g_1+g_2) + T_W$, one can evaluate the ensemble integrals analytically. This 
yields
\begin{equation}
\fl
\hat C[\sigma_\tot^{(a)},\sigma_\tot^{(b)}](t) = T_A^2\; \rme^{-T_W\, t} \left\{
   (1+2\delta_{ab})\; (1+2 T_A\, t)^{-3} - (1+T_A\, t)^{-4}\; b_2(t) \right\} \; .
\label{BreitWigner}\end{equation}

In the GOE case, expression~(\ref{Wabsorp}) can also be verified for the VWZ integral
\cite{Ver85a}, which gives an analytical expression for the exact correlation function
in equation~(\ref{Csigtot}). The Fourier transform of the VWZ integral reads \cite{Gor02}:
\begin{equation}
\fl C[S_{ab},S_{cd}^*](t) = \frac{1}{4}\int_{\max(0,t-1)}^t \rmd r \;
(t-r) (r+1-t) \prod_{e=1}^M\left[1-T_e(t-r)\right] \; U(r) \; ,
\label{VWZint}\end{equation}
where
\begin{eqnarray}
\fl
U(r) & = & 2\int_0^{r^2}\rmd x \;
\frac{\delta_{ab}\delta_{cd}\; \Delta_a \Delta_c
\; + \; (\delta_{ac}\delta_{bd}+\delta_{ad}\delta_{bc}) \; \Pi_{ab}}
{(t^2-r^2+x)^2 \; \sqrt{x(x+2r+1)} \; \sqrt{\prod_{e=1}^M(1+2T_e r+T_e^2 x)}}
\; , \label{V_U1}\\
\fl \Delta_a & = & 2T_a \sqrt{1-T_a} \left( \frac{r+T_a x}{1+T_a(2r+T_a x)} +
\frac{t-r}{1-T_a(t-r)} \right) \; , \label{V_Delta}\\
\fl \Pi_{ab} & = & 2 T_a T_b \left(
\frac{T_a T_b x^2 + [T_a T_b r + (T_a+T_b)(r+1)-1] x + r(2r+1)}
{(1+2T_a r+T_a^2 x)(1+2T_b r+T_b^2 x)} \right. \nonumber\\
\fl && + \left. \frac{(t-r)(r+1-t)}{[1-T_a(t-r)] \; [1-T_b(t-r)]} \right) \; .
\end{eqnarray}
We split the products occurring in expressions~(\ref{VWZint}) and
(\ref{V_U1}) and consider that part running over the absorption channels.
In the asymptotic limit of equation~(\ref{Wabsorp}), we find
\begin{eqnarray}
\prod_{c=1}^{M_W} \left[1 - T_{M_A+c} (t-r)\right] \to \rme^{-T_W\, (t-r)} \\
\prod_{c=1}^{M_W} [1+2\, T_{M_A+c}\, r + T_{M_A+c}^2\, x]^{-1/2}
    \to \rme^{-T_W\, r} \; ,
\end{eqnarray}
The $r$-dependent exponentials cancel, which proves our conjecture for the
GOE case.

\begin{figure}
\begin{center}
\includegraphics[width=0.48\textwidth]{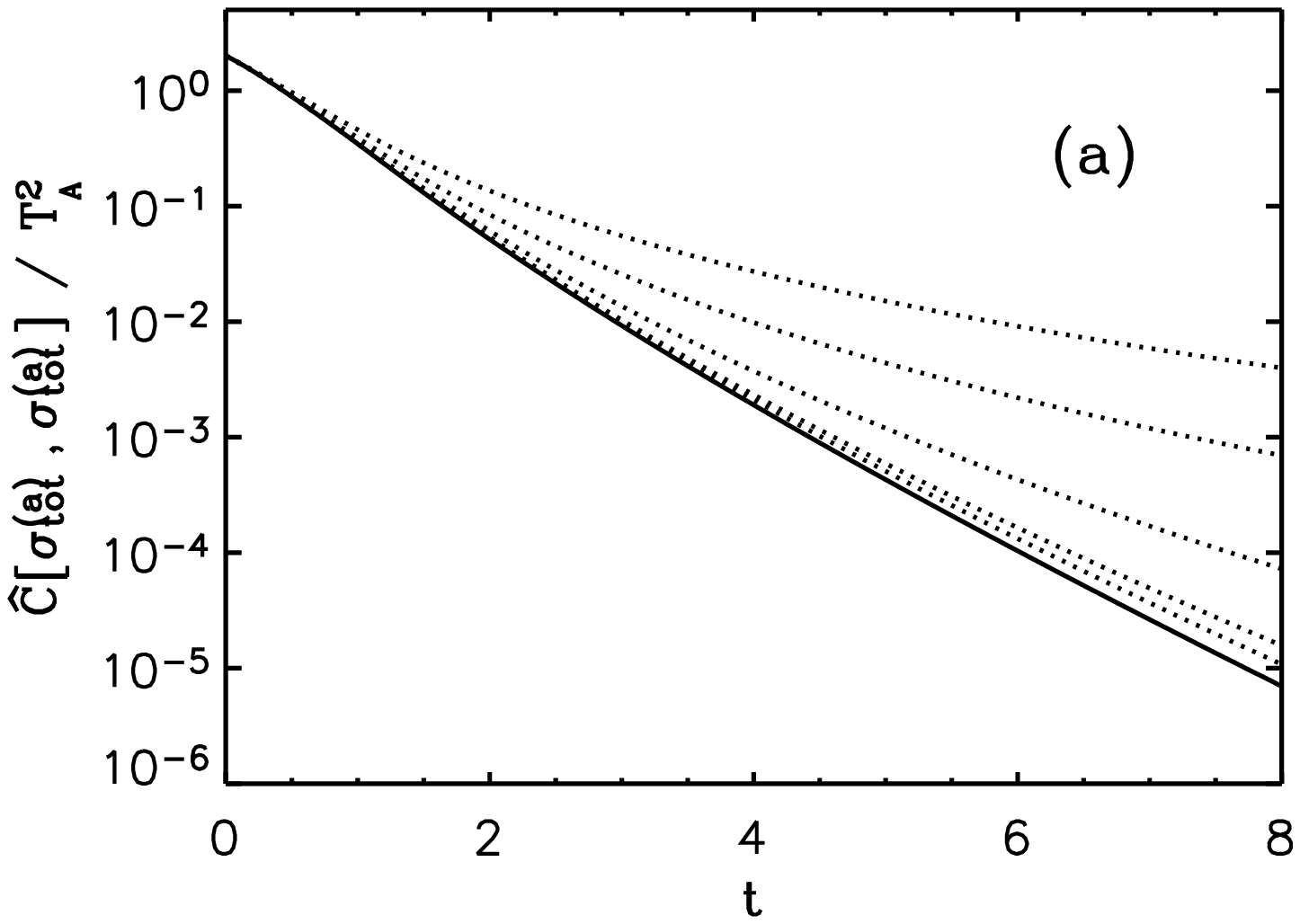}
\includegraphics[width=0.48\textwidth]{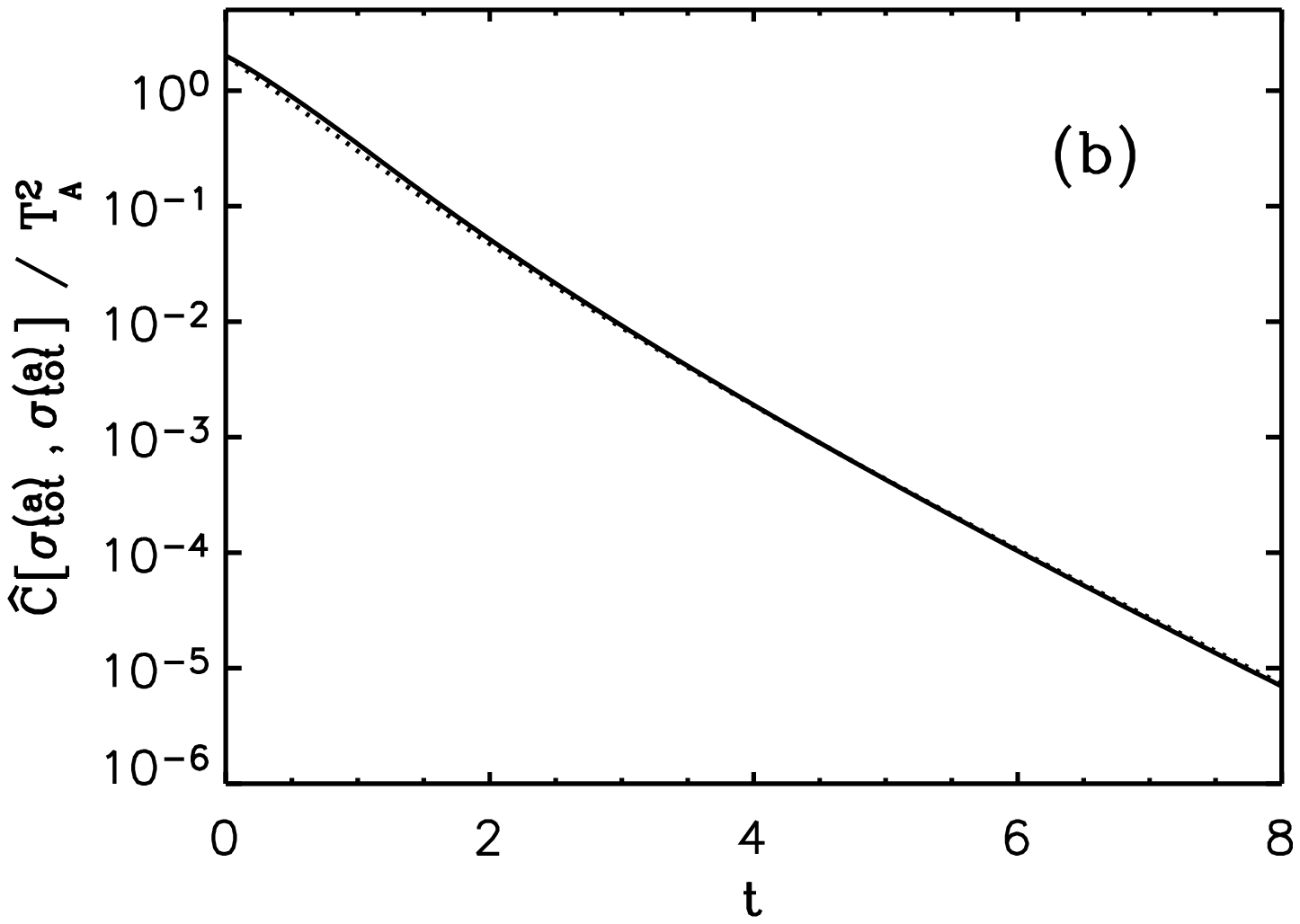}
\end{center}
\caption{ Auto-correlation
functions for the GOE case with two accessible channels (antennas) with equal
transmission coefficients $T_A = 0.26$ and total transmission for the wall
absorption: $T_W = 1$, where the wall absorption is equally distributed over $M_W$ channels. In (a) the VWZ result is shown as dotted lines for
$M_W= 1, 4, 16, 64, 128$, and as a solid line for
$M_W=\infty$. 
In (b) we compare the exact VWZ result (solid line) with the  rescaled Breit-Wigner result (dotted line) for $M_W=\infty$.}
\label{F_Mwand}\end{figure}

It will be useful to investigate the effect of the number of absorption channels
$M_W$ on the correlation function. For this purpose, we assume the total wall 
absorption to be fixed, $T_W= 1$, and distributed equally among the absorption
channels $T_{M_A+c} = T_W/M_W$. Figure~\ref{F_Mwand}(a) shows the 
behavior of the auto-correlation function for the case of two antennas with 
$T_A=T_1=T_2=0.26$ for different values of $M_W$. Note the significant 
changes in the shapes of the curves when $M_W$ is varied. 
There are two previous microwave experiments where the channel number dependence of the auto-correlation function was studied. The first one by Doron et al. \cite{Dor90} failed to see any difference between exponential and algebraic decay behavior \cite{Lew92}. However, in the more recent work by Alt et al. \cite{Alt95a} an algebraic decay of the auto-correlation function was observed.

In figure \ref{F_Mwand}(b) we compare the exact result with the result from the rescaled
Breit-Wigner approximation for $M_W=\infty$ using the same values for $T_A$ and $T_W$ as in figure~\ref{F_Mwand}(a).
The difference between the RBWA and the exact result is very small. This permits us to use the much simpler and more flexible RBWA in the analysis of our experimental data.
For larger values of $T_A$ the RBWA is less accurate than in the example above, but it can still be used to determine $T_W$ reliably from the auto-correlation function. The accuracy of the RBWA is discussed in detail in reference~\cite{Gor02}. 

\section{Experiment} \label{sec:experiment}

Since the experiment is described in detail elsewhere \cite{Kuh00b}, we concentrate
on the aspects relevant in the present context. Reflection and transmission
measurements have been performed in microwave cavities of various shapes.
All cavities are flat, with top and bottom plate parallel
to each other. The cavities are quasi-two-dimensional for frequencies
$\nu < \nu_{max}=c/(2h)$ ($h$: height of the billiard). In this regime there is a complete
equivalence between the stationary wave equation and the corresponding stationary
Schr\"odinger equation, where the $z$ component of the electric field corresponds to the quantum
mechanical wave function,
\begin{equation}
\Delta\; \Psi(x,y) + E\; \Psi(x,y) = 0 \; , \quad
E = \left(\frac{2\pi\, \nu}{c}\right)^2 \; ,
\label{eq:Enu}\end{equation}
with Dirichlet boundary conditions.
The antennas consisted of copper wires with a diameter of $1\,$mm, projecting
$l_p=2$ or $4\,$mm into the resonator. An Agilent 8720ES vector network analyzer was
used to determine the complete S-matrix. Measurements were taken in
the frequency range from $1$ to $16\,$GHz with a resolution of $0.5\,$MHz.

The unwanted contribution of the cables to the S-matrix was removed by
standard calibration procedures. It was not possible, however, to get rid of
the contribution of connectors and antennas in this way. This posed a
problem in particular for the reliable determination of the phase, which is
vital for the cross correlation measurements. Therefore, the phase shift from
the antennas was determined from a reference measurement, where the cavity was
removed, and only the antennas and the supporting top plate were present. We
checked that the average S-matrix is in good approximation real and diagonal.

\begin{table}
  \centering
\begin{tabular}{|c|c|c|r|r|r|r|}\hline
  billiard type & shape & material & $A$/cm$^2$ & $L$/cm & $h$/mm & $l_p$/mm
  \\ \hline \hline \hline
  \parbox{16ex}{rectangle \\ $34 \times 24$\,cm$^2$} & 
     \parbox{2cm}{\includegraphics[width=2cm]{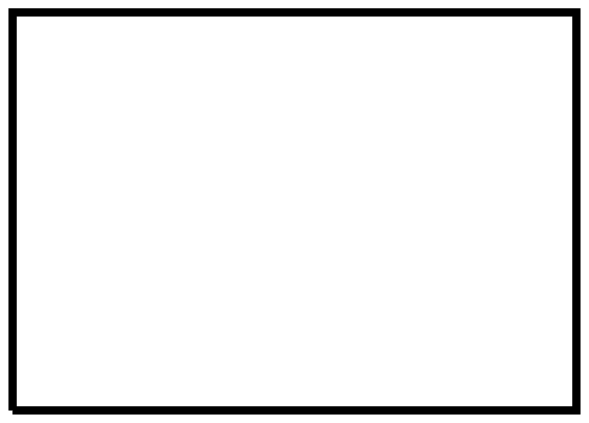}} & 
     brass & 816 \ & 116.0 & 8 \ \ & 2 \ \ \\ \hline
  \parbox{16ex}{Robnik billiard \\ $\lambda=0.4$} &
     \parbox{2cm}{\includegraphics[width=2cm]{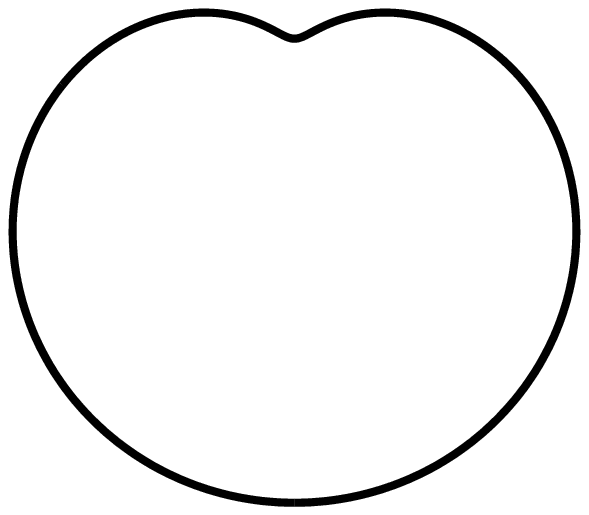}} & 
     brass & 474 \ & 77.5 & 8 \ \ & 2 \ \ \\ \hline
  \parbox{16ex}{billiard with \\ $C_3$ symmetry} &
     \parbox{2cm}{\includegraphics[width=2cm]{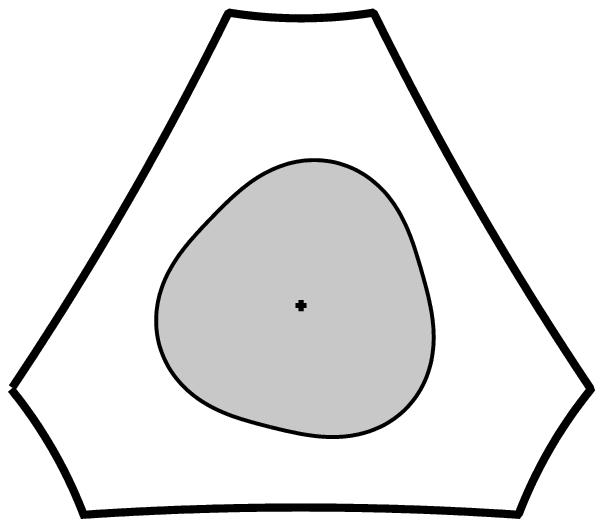}} & 
     aluminum & 2580 \ & 358.5 & 10 \ \ & 4 \ \ \\ \hline
  \parbox{16ex}{fully chaotic \\ billiard} & 
     \parbox{2cm}{\includegraphics[width=2cm]{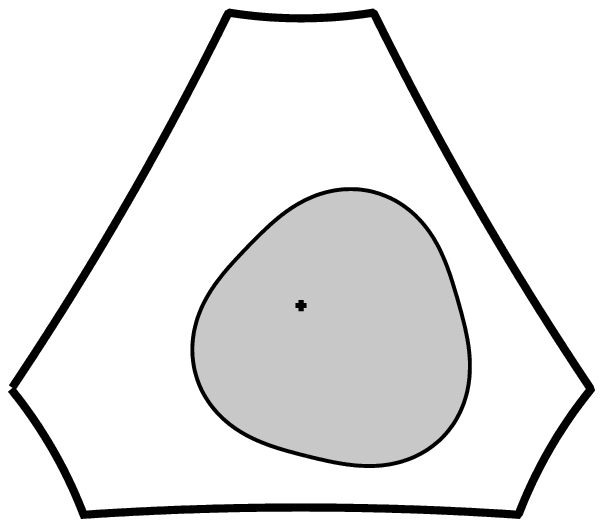}} & 
     aluminum & 2580 \ & 358.5 & 10 \ \ & 4 \ \ \\ \hline 
\end{tabular}
  \caption{Details of the studied billiards, with $A$: area, $L$: circumference, $h$: height of the billiard, and $l_p$: the projection length of the antennas into the billiards.}\label{tab:overview}
\end{table}

Four different cavities were used, which are presented in
table \ref{tab:overview}. 
The rectangular and the Robnik billiard both have fixed geometries. Therefore,
no ensemble average can be taken for these systems. Instead, we performed 10 measurements with different antenna positions on each system. By this we
obtained 10 spectra of essentially the same system (it is slightly
altered by the change of the antenna positions), but
with different intensities for each of the resonances.

The billiard with threefold symmetry has been discussed in detail in
\cite{Sch02}. It is composed of an outer part  and an insert, both with $C_{3v}$
symmetry. By rotating the insert we get an ensemble of systems all displaying
$C_3$ symmetry. However, configurations with $C_{3v}$ symmetry have been
avoided. We performed 30 measurements with different rotation angles and
fixed antenna positions. The two antenna positions were in symmetry
equivalent or non-equivalent positions, alternatively.

The fully chaotic billiard is a variant of the $C_3$ billiard,
where the insert was placed out of the center avoiding any symmetry.
We performed 50 measurements for different positions of the insert.
The classical dynamics for the latter two billiards is completely chaotic.

\begin{figure}
\centerline{ \includegraphics[width=0.48\textwidth]{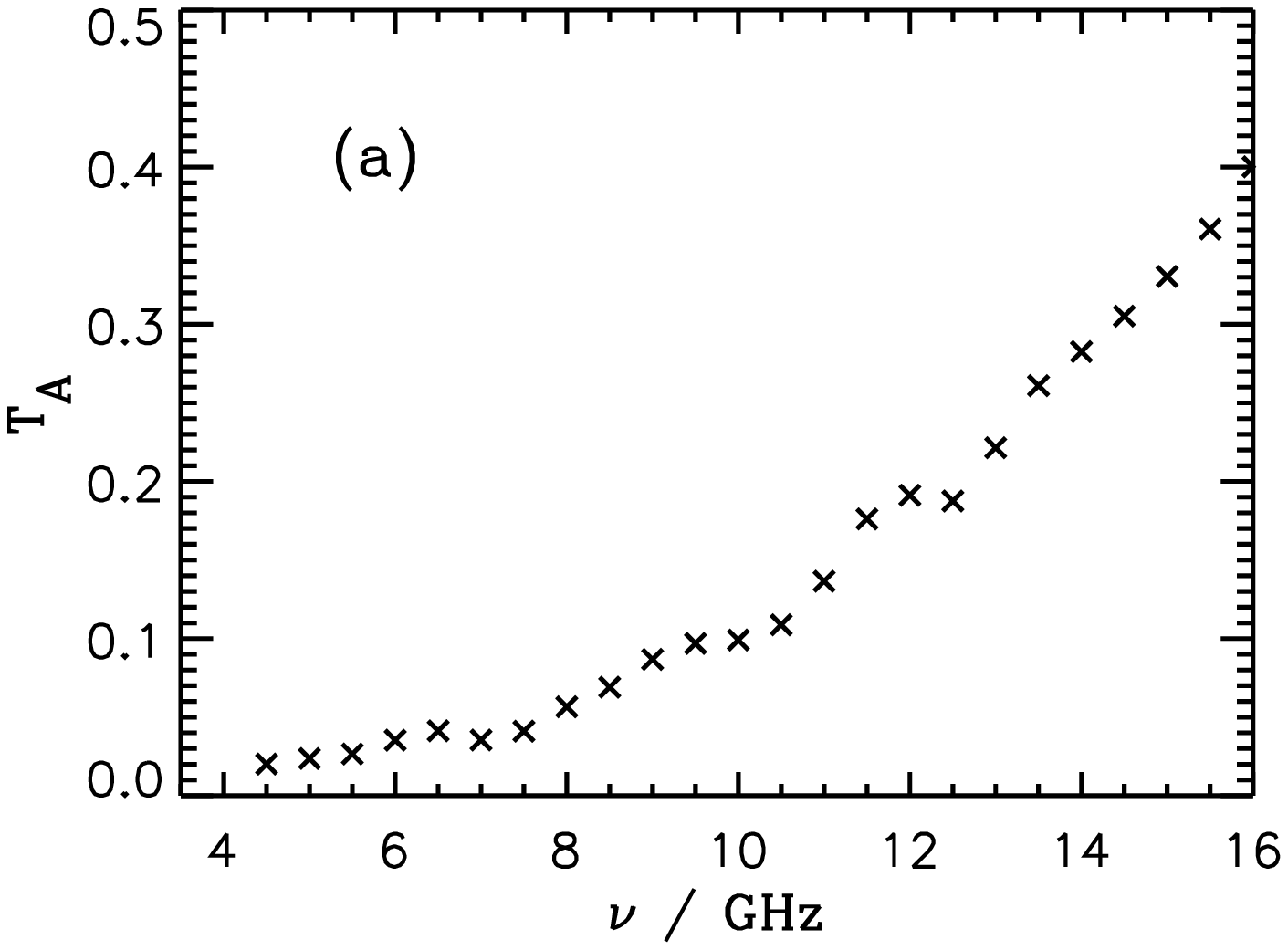} 
 \includegraphics[width=0.48\textwidth]{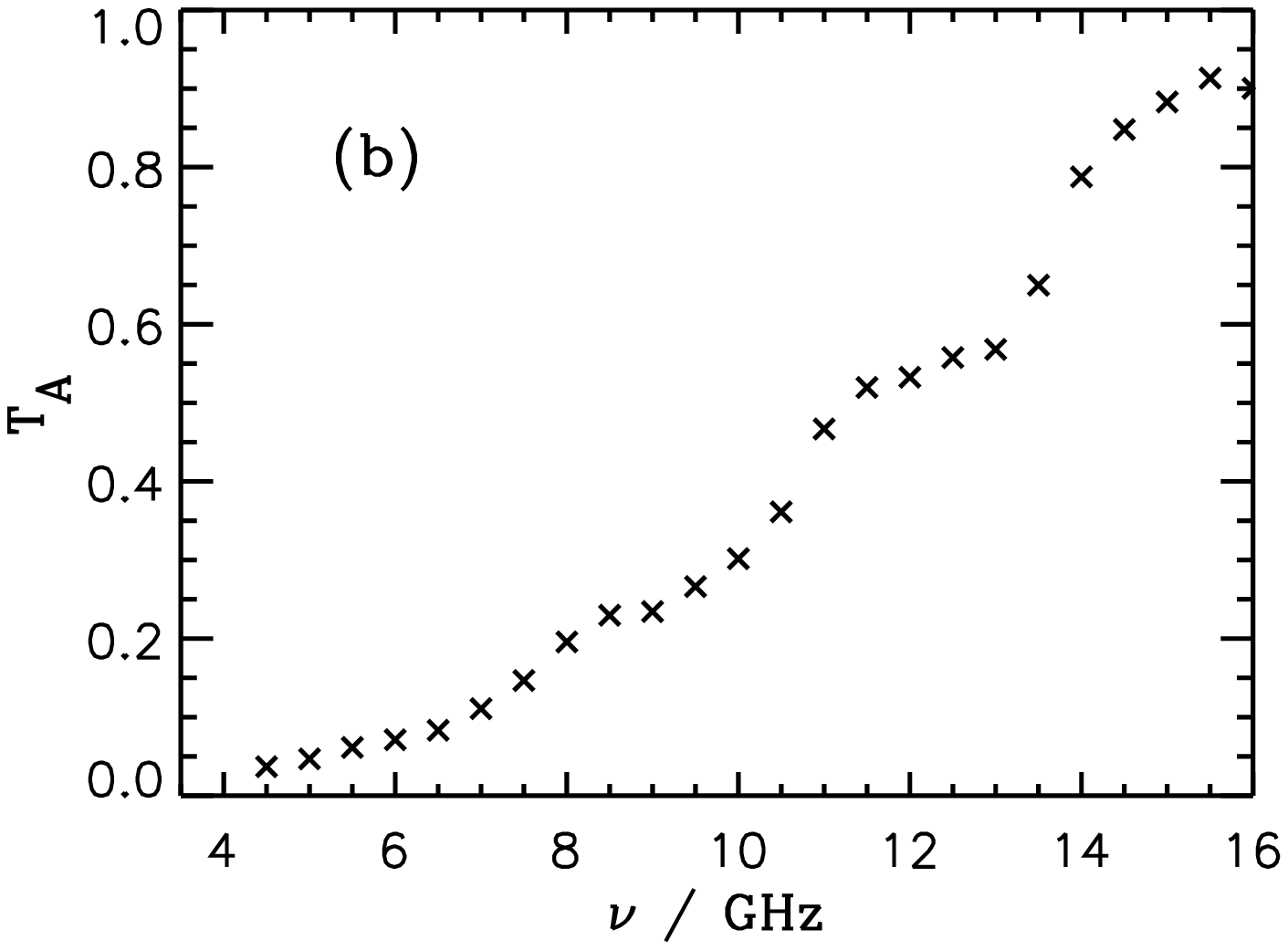} }
\caption{Transmission coefficient $T_A$ obtained from the average
S-matrix for (a) the rectangular and (b) the fully chaotic billiard.}
\label{fig:Tant}
\end{figure}

As described in section~\ref{B}, the transmission coefficients $T_A$ of the
antennas are obtained from the average S-matrix (see equations~(\ref{eq:kappa})
and (\ref{defT})). In figure~\ref{fig:Tant} the results for $T_A$ as a
function of frequency are shown for the rectangular and the fully chaotic billiard. 
One notices a strong frequency dependence of the coupling. Such a behavior is typical for wire antennas. For each system, the two antennas yield approximately the same transmission coefficients.
In view of the frequency dependence of $T_A$ and a comparable one of $T_W$
to be discussed in section~\ref{sec:autocor}, we examined frequency intervals with a width of 1\,GHz to assure that the average S-matrix and the total absorption are
approximately constant.

The frequency dependence of the wall absorption is mainly due to the skin effect~\cite{Jac62}.
The additional width acquired by the resonances is
\begin{equation}
\Gamma_W(\alpha)= \frac{8\pi^2}{c^2}\; \nu^2\; \frac{\delta(\nu)}{h}\left(
   1+ \frac{h\, L}{2A}\; \xi(\alpha)\right) \; , \quad
\delta(\nu)= \frac{1}{\sqrt{\pi\mu_0\sigma\nu}} \; ,
\end{equation}
where $\delta(\nu)$ is the so-called skin depth, and $\mu_0$ and $\sigma$ are the
susceptibility and the conductance of the cavity walls, respectively. The
dimensionless quantity $\xi(\alpha)$ is of order one. It depends on the particular
resonances, indexed by $\alpha$.

To describe the wall absorption within our statistical model, we consider the
average $\bar\Gamma_W$ and the variance $(\Delta\Gamma_w)^2$ of the absorption width.
The number of absorption channels $M_W^\eff$ (with equal transmission coefficients)
can be determined taking into account that in the statistical model $\Gamma_W$ is the
sum of $M_W^\eff$ random Porter-Thomas variables. In order to reproduce the average
absorption width and its variance, it must hold:
\begin{equation}
\frac{1}{M_W^\eff} = \frac{(\Delta\Gamma_W)^2}{\bar\Gamma_W^2} 
   \approx \left(\frac{h\, L}{2A}\right)^2\; (\Delta\xi)^2 \; ,
\end{equation}
where $(\Delta\xi)^2$ is the variance of $\xi(\alpha)$ as it fluctuates for different
resonances. As $\xi(\alpha)$ is of order one, its variance cannot be larger. Therefore,
$M_W^\eff \gtrsim [2A/(hL)]^2$, which is greater than $200$ for the studied cavities.
This is certainly indistinguishable from an infinite number of channels.

\section{\label{sec:autocor}Auto-correlation function}

\begin{figure}
 \centerline{ \includegraphics[width=0.48\textwidth]{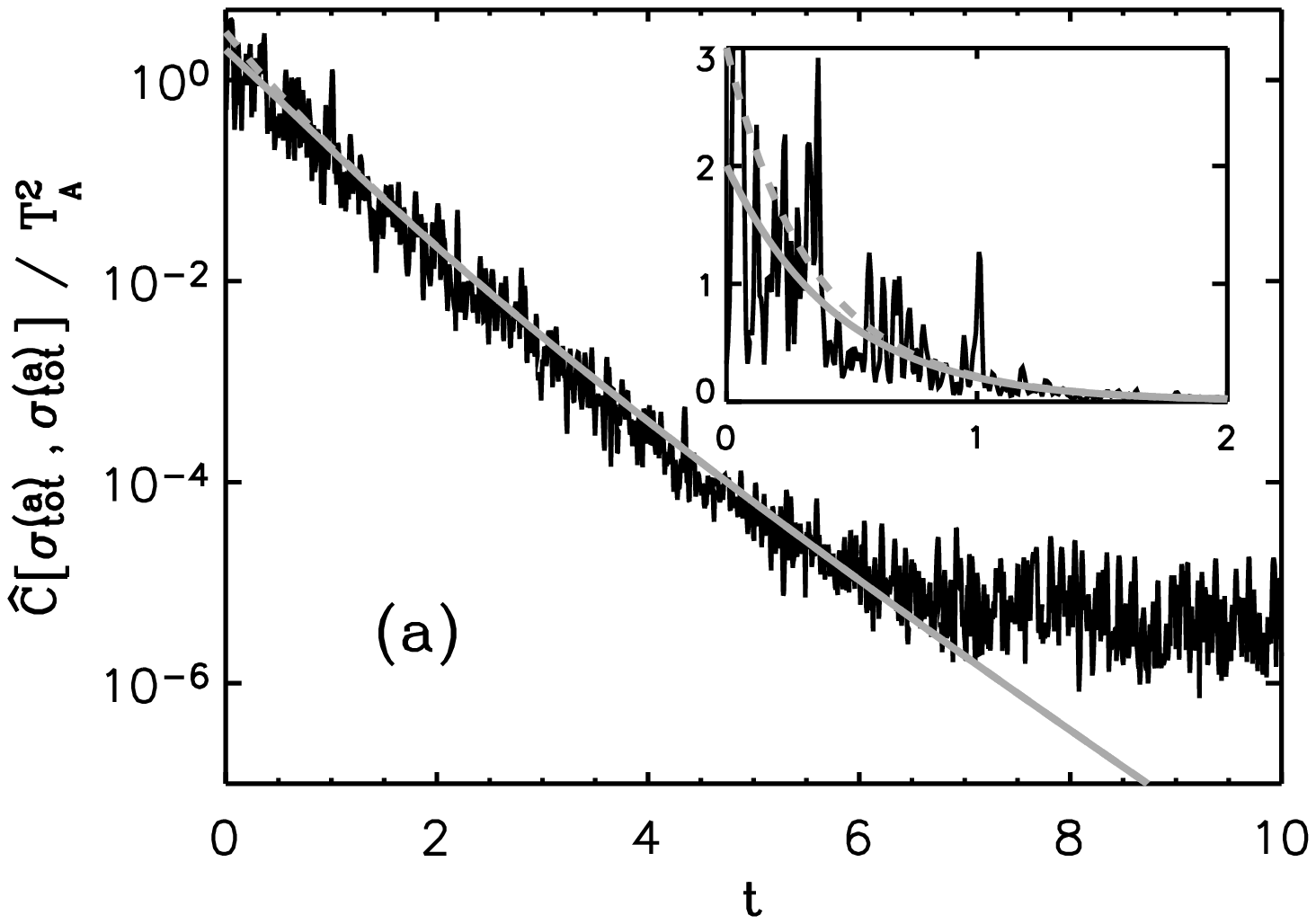}
\includegraphics[width=0.48\textwidth]{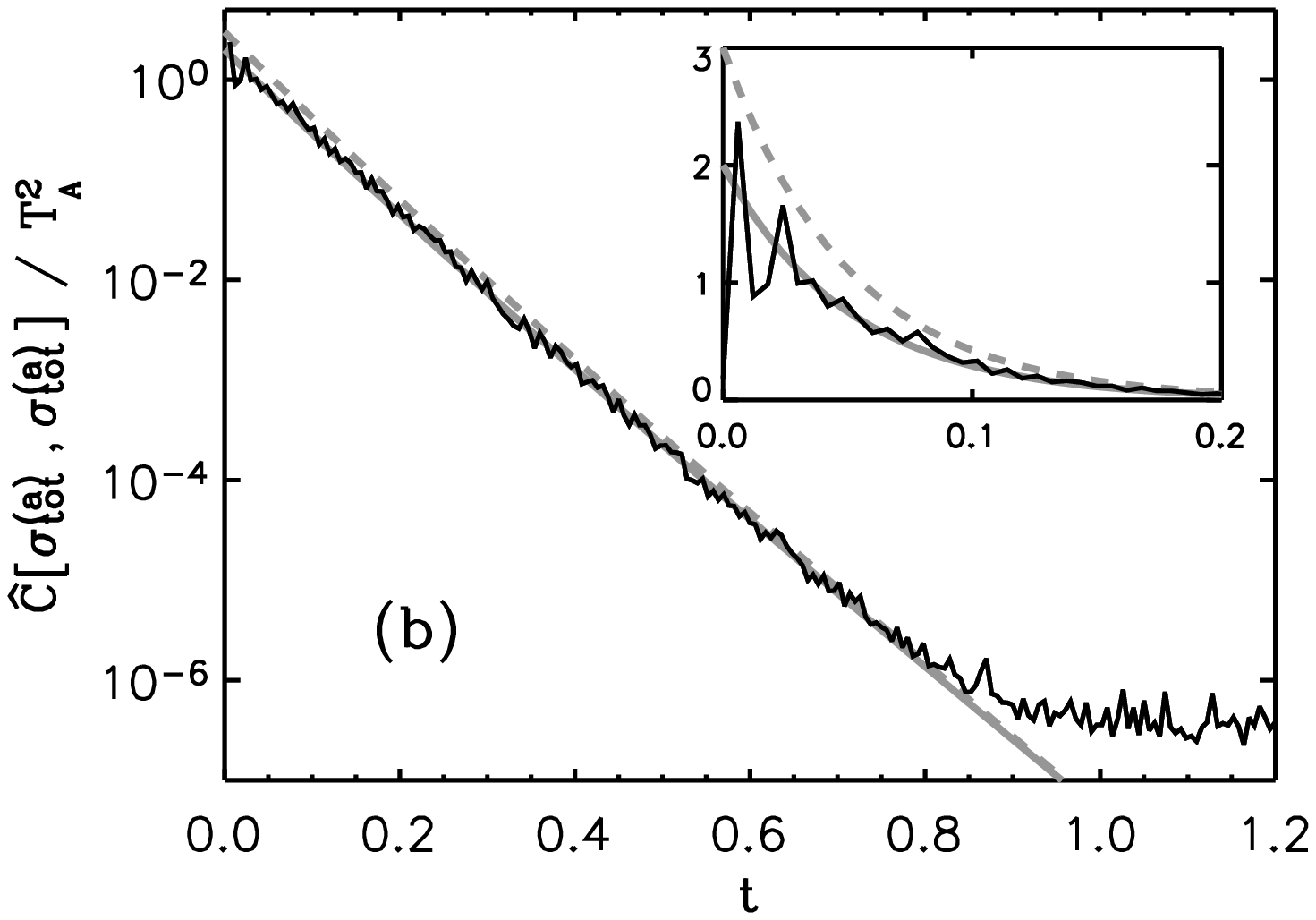}}
 \caption{Logarithmic plot of $\hat C[\sigma_{tot}^{(a)},\sigma_{tot}^{(a)}] / T_A^2$ 
for (a) the rectangular and (b) the fully chaotic billiard obtained from Fourier
transforms over the frequency ranges 13 to 14\,GHz and 14 to 15\,GHz,
respectively. The insets show the corresponding linear plots for small times $t$. 
The theoretical prediction from the RBWA is shown
for the GOE (solid line) and the POE (dashed line).
The parameters were $T_A=0.261$, $T_W=1.38$ for the rectangle, and
$T_A=0.848$, $T_W=14.92$ for the fully chaotic billiard.}
 \label{fig:acor}
 \end{figure}

In this section we examine the Fourier transform of the
auto-correlation function $\hat C[\sigma_\tot^{(a)},\sigma_\tot^{(a)}](t)$ as given in
equation~(\ref{defCtot}).
Figure \ref{fig:acor} shows logarithmic plots of the auto-correlation functions
for the rectangular and the fully chaotic billiard together with the rescaled Breit-Wigner
approximation for GOE and POE - assuming an infinite number of weakly coupled
channels for the absorption in the walls (see equation~(\ref{BreitWigner})).
The results for the rectangular billiard do not allow us to distinguish between GOE and POE behavior due to their large fluctuations. At most we can see a hint of the fact that the auto-correlation function for rectangular billiards tends to 2.25 instead of 3 as $t \to 0$. 
For rectangular billiards the squared amplitudes (entering into equation~(\ref{CstRBW})) are not Porter-Thomas distributed, leading to an auto-correlation function that is closer to the prediction for GOE than to the one for POE. 

The ensemble averaged auto-correlation function of the fully chaotic billiards shows much smaller fluctuations, of course, and we observe a very good agreement with the GOE prediction. The correlation hole can be seen in particular in the linear plot shown in the inset of figure~\ref{fig:acor}(b). 
However, the correlation hole is reduced to $1/3$ of its full value (see  section~\ref{sec:intro}), and the difference to integrable systems may be even smaller.

The experimental auto-correlation function follows the theoretical
 curve for an infinite number of absorption channels over five orders of
 magnitude. Then the auto correlation deviates from the theoretical curve. 
This is a consequence of the finite frequency interval used in the Fourier transform. The long-time behavior is dominated by the Welch filter applied.
Comparison with figure \ref{F_Mwand} shows that more than 100 weakly coupled
channels have to be assumed to explain this behavior, but the simplifying
assumption of
infinitely many channels is in accordance with the experiment as well.
Further, we observe that the rescaled Breit-Wigner approximation is
sufficient to describe the experimental results.

\begin{figure}
\centerline{ \includegraphics[width=0.48\textwidth]{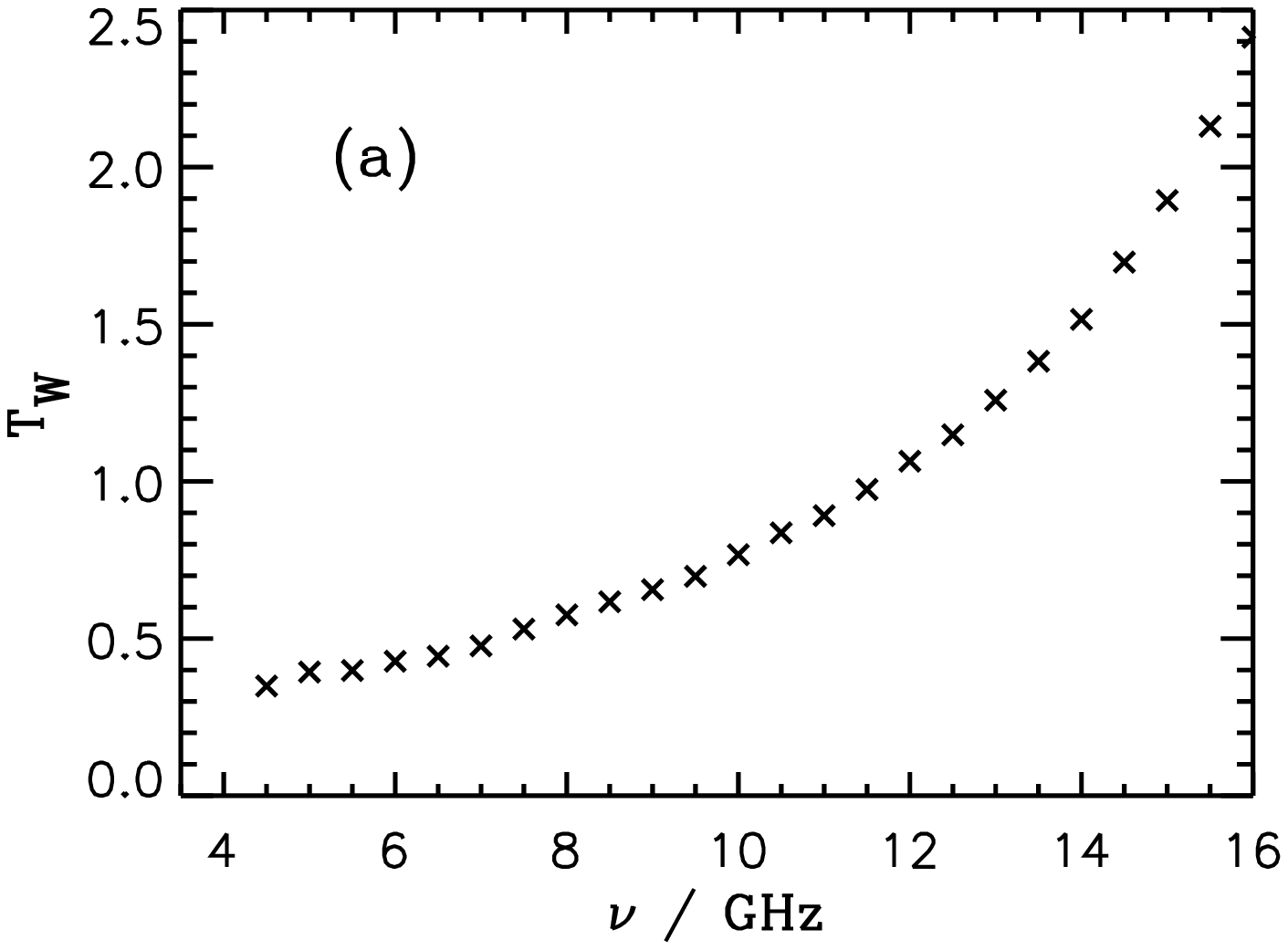}
\includegraphics[width=0.48\textwidth]{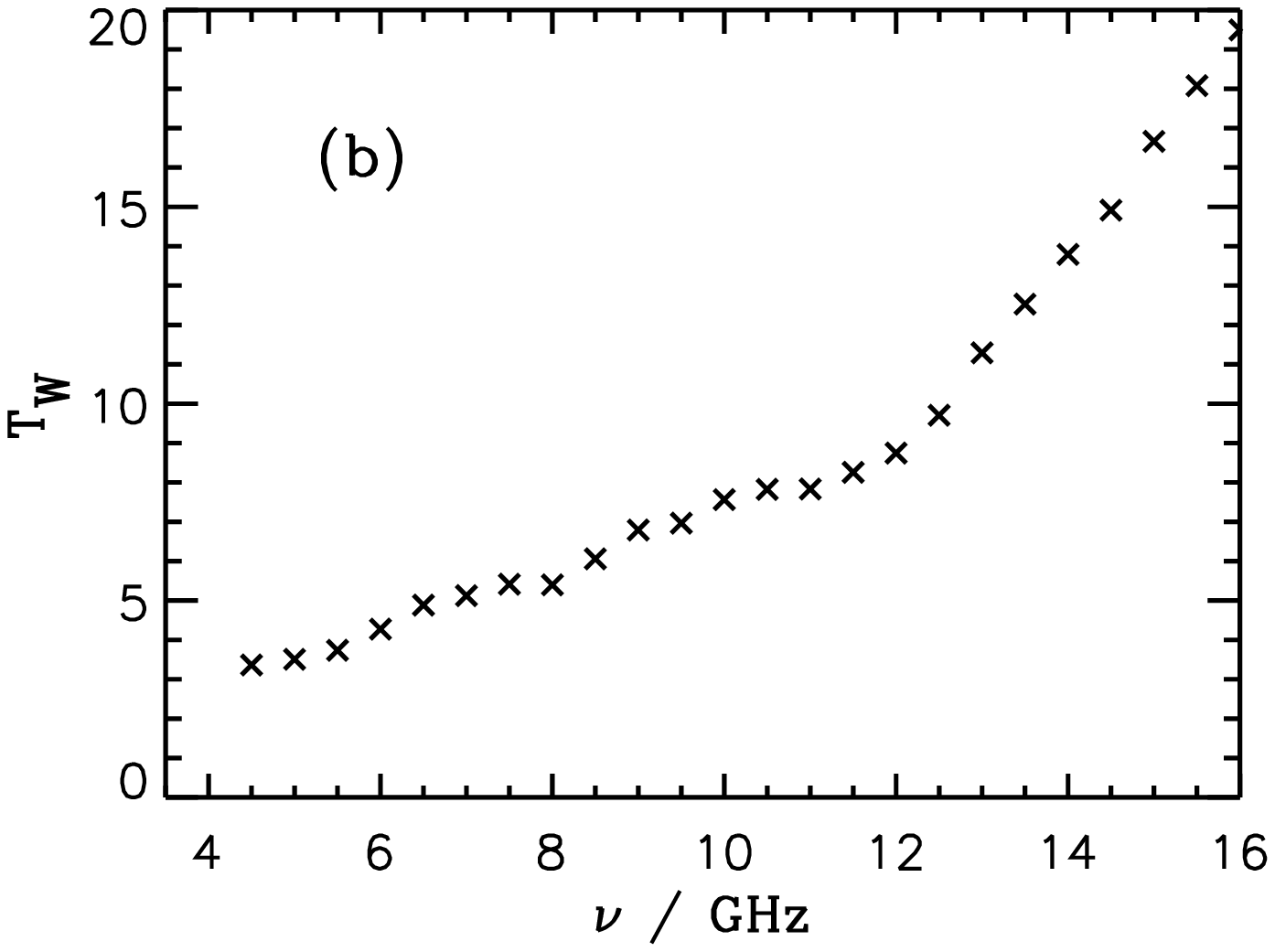} }
\caption{Transmission $T_W$ into the walls, obtained from a fit to the auto-correlation function for
(a) the rectangular and (b) the fully chaotic billiard.}
\label{fig:Twand}
\end{figure}

As the antenna transmission $T_A$ has been obtained independently, the wall
transmission $T_W$ can be determined
by fitting the experimental auto-correlation function with the corresponding rescaled
Breit-Wigner expression.
This procedure works well for a large frequency range, and all results
presented in this paper have been obtained in this way. In the low frequency
regime, however, a two-parameter fit of the auto-correlation function, treating
both $T_W$ and $T_A$ as free parameters, yielded somewhat better results.
Figure \ref{fig:Twand} shows the frequency dependence of $T_W$, as determined from the auto-correlation function, both for the rectangular and for the fully chaotic billiard.

\section{\label{sec:cross}Cross-correlation function}

\begin{figure}
\centerline{ \includegraphics[width=0.48\textwidth]{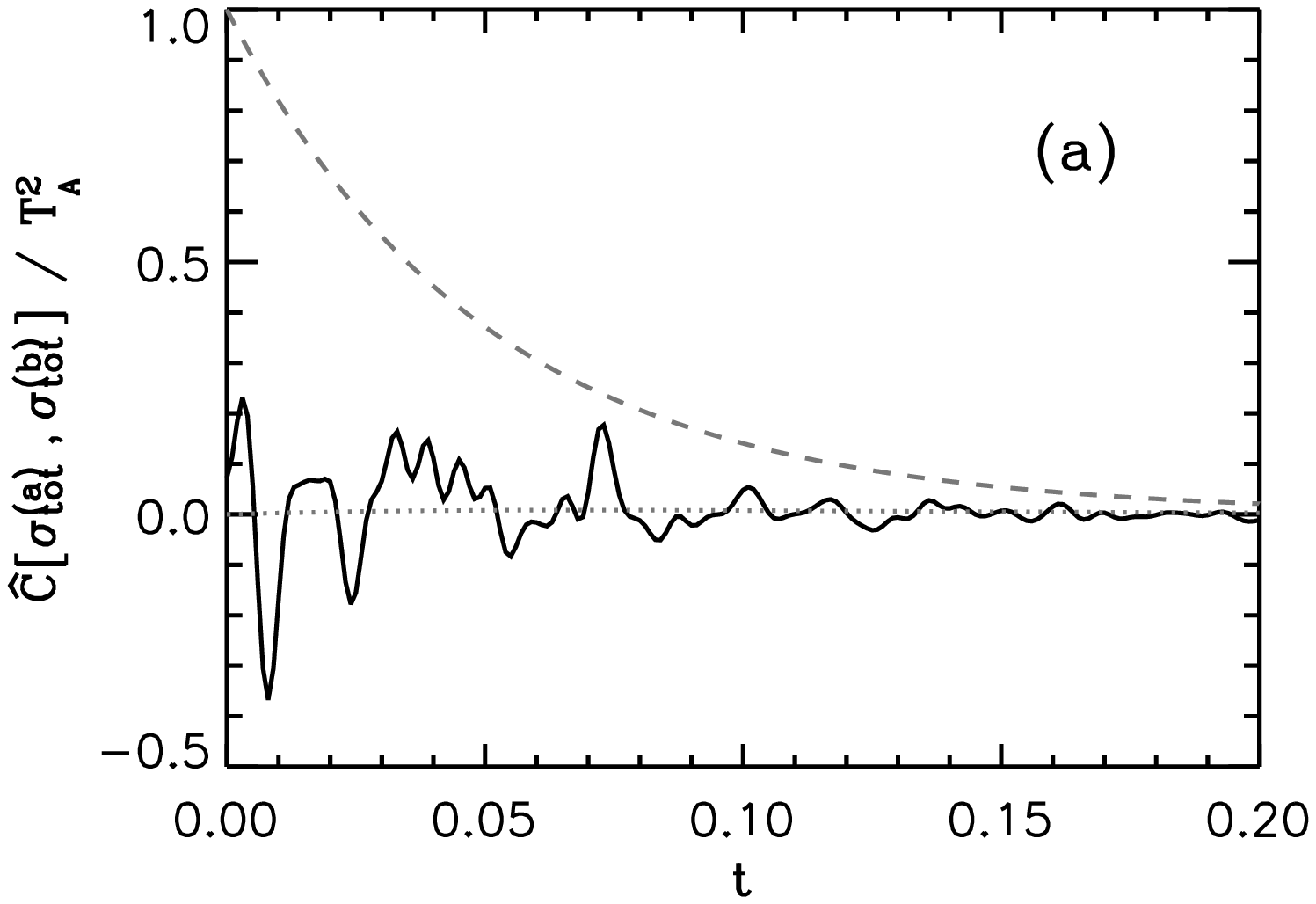}
\includegraphics[width=0.48\textwidth]{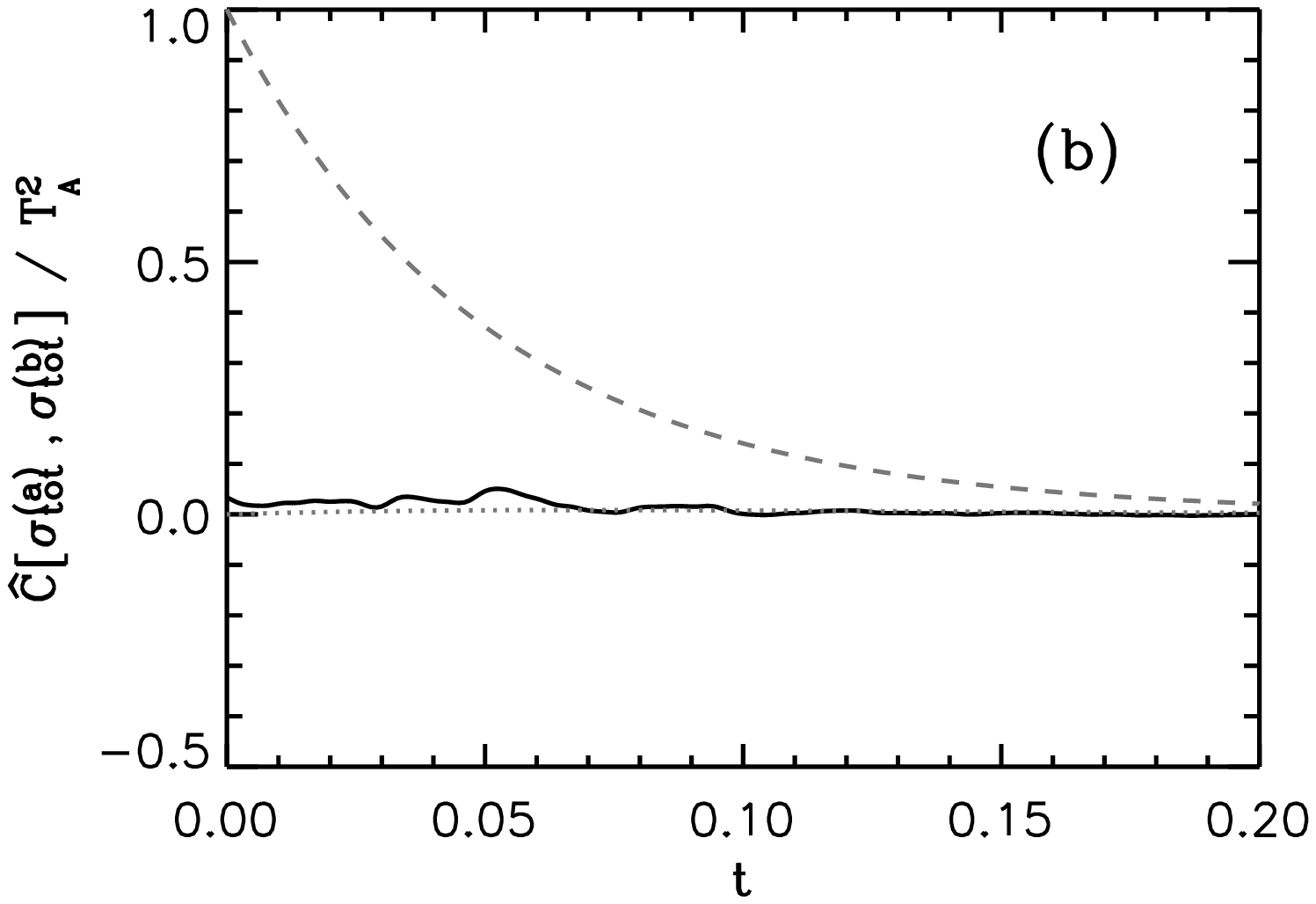}}
\caption{Cross-correlation function $\hat C[\sigma_{tot}^{(a)},\sigma_{tot}^{(b)}] / T_A^2$ for the fully chaotic billiard, averaged
over 50 realizations, (a) before smoothing and (b) after smoothing.
The Fourier transform was taken over the frequency range of 14 to 15\,GHz.
The RBWA is shown for the GOE (dotted line) and for the POE (dashed line)
with $T_A=0.848$ and $T_W=14.92$.}
\label{fig:xcor}
\end{figure}

We now present cross-correlation functions $\hat C[\sigma_\tot^{(a)},\sigma_\tot^{(b)}](t)$
of the total cross sections, which are equivalent to the ones of the respective
diagonal S-matrix elements, see equation (\ref{Csigtot}).
Figure~\ref{fig:xcor} shows $\hat C [\sigma_{tot}^{(a)},\sigma_{tot}^{(b)}] / T_A^2$ for the fully
chaotic billiard. The
dotted and the dashed lines correspond to the expectation from the
rescaled Breit-Wigner approximation.

Even after averaging over 50 realizations, the fluctuations are still quite
strong. Therefore we apply a smoothing over an
interval of size $\ln 2/T_W$, leading to the smooth behavior displayed in
figure~\ref{fig:xcor}(b). All results presented below are smoothed in this way.

\begin{figure}
 \centerline{ \includegraphics[width=0.48\textwidth]{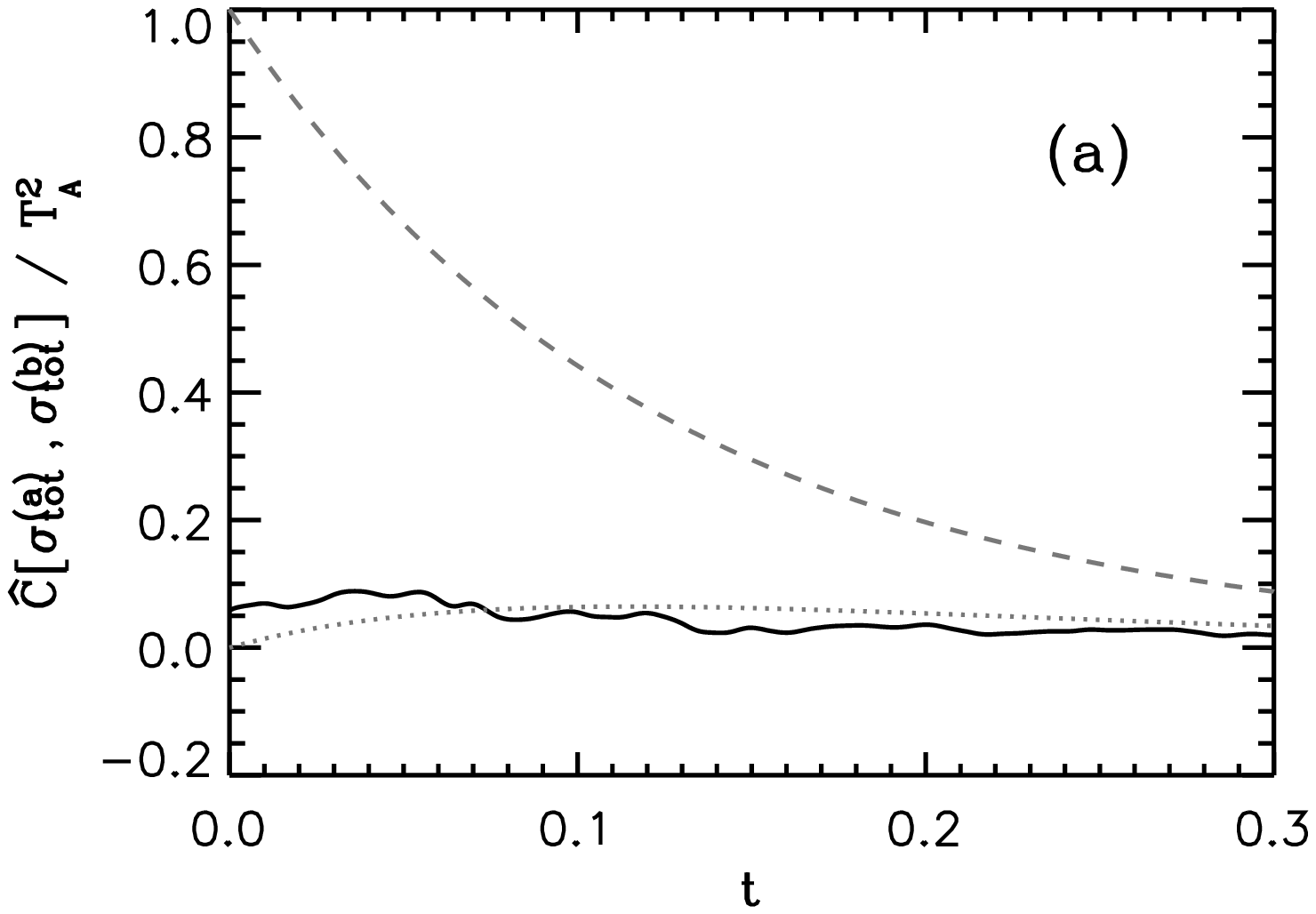}
  \includegraphics[width=0.48\textwidth]{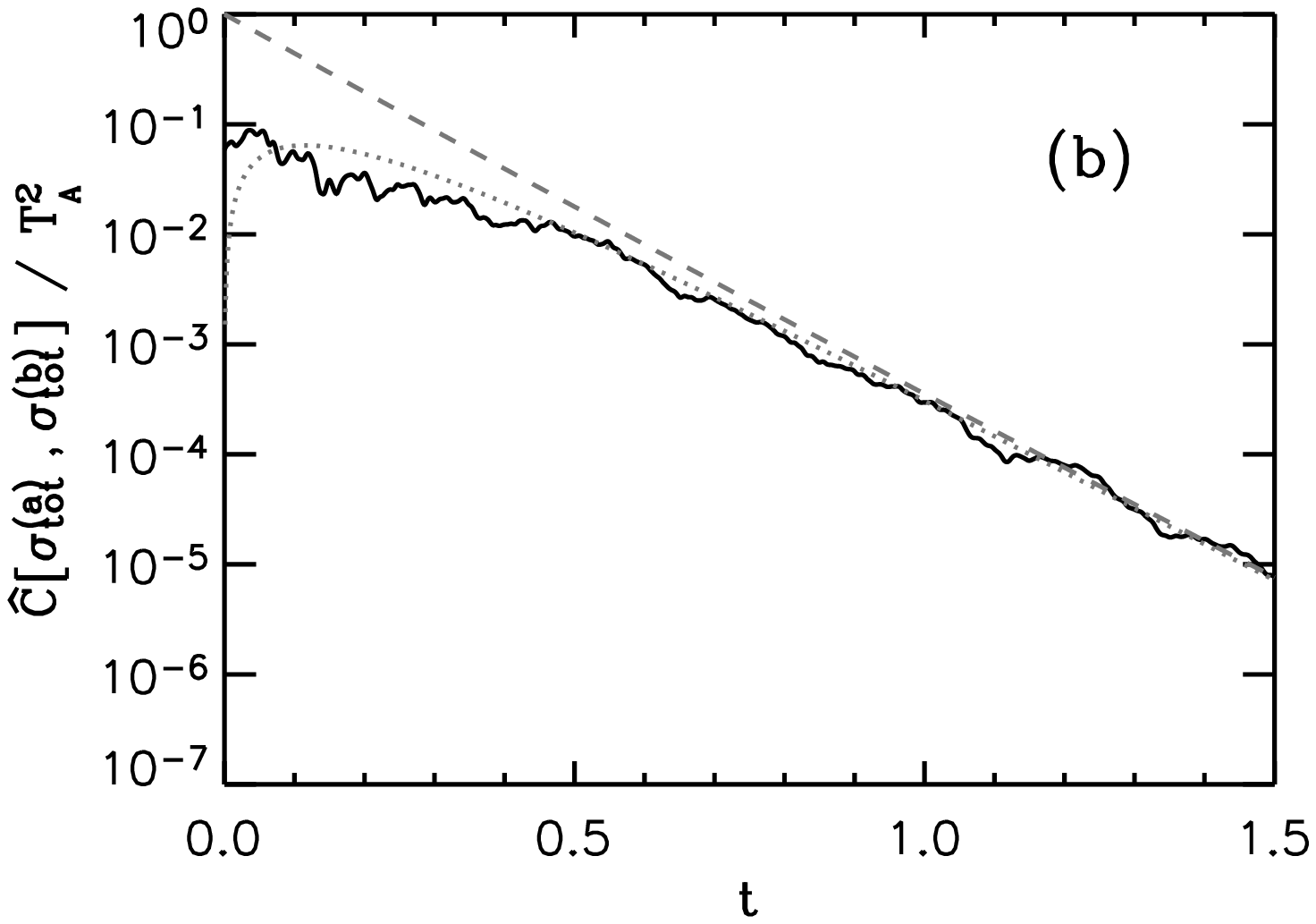} }
 \centerline{ \includegraphics[width=0.48\textwidth]{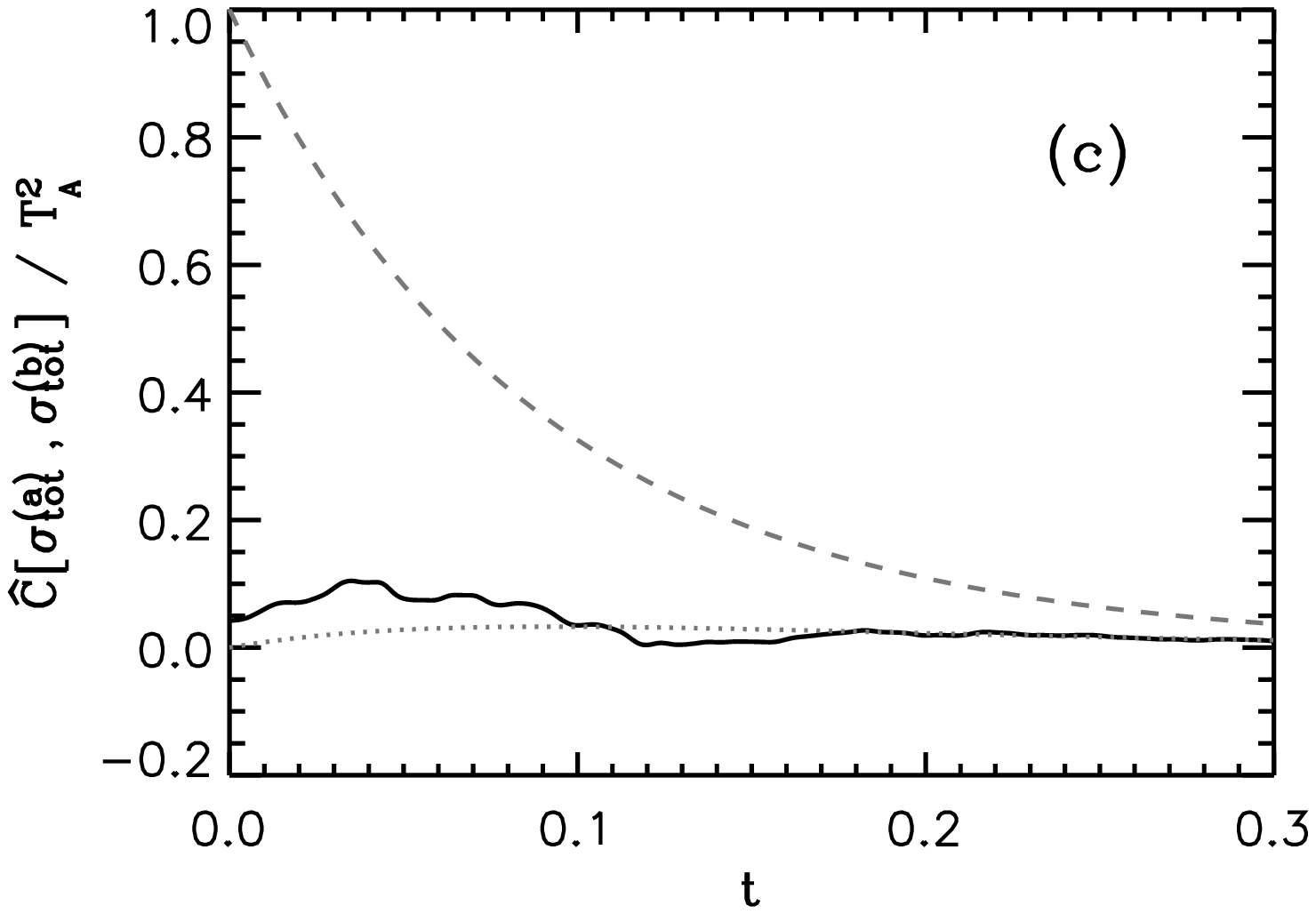}
  \includegraphics[width=0.48\textwidth]{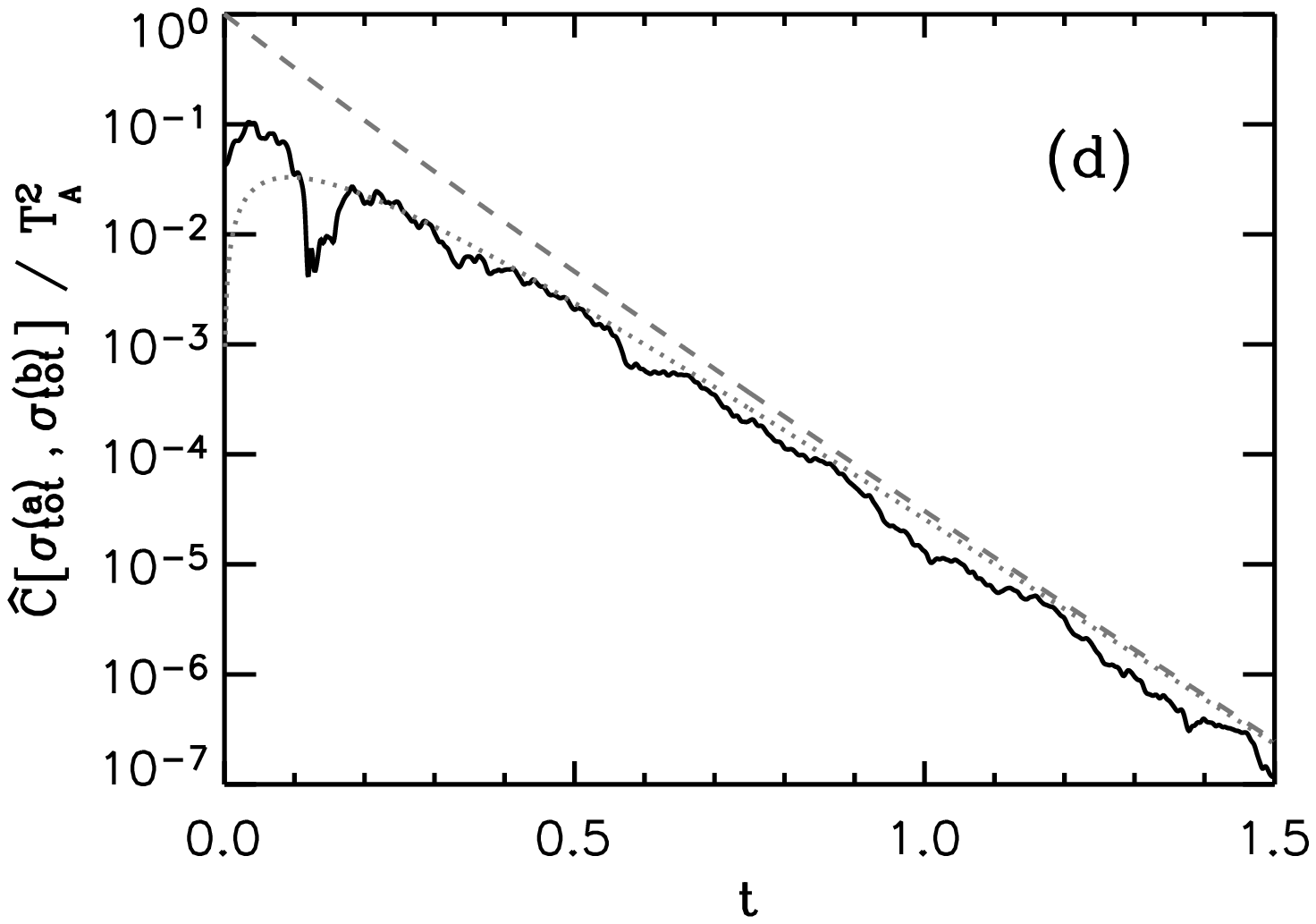} }
 \centerline{ \includegraphics[width=0.48\textwidth]{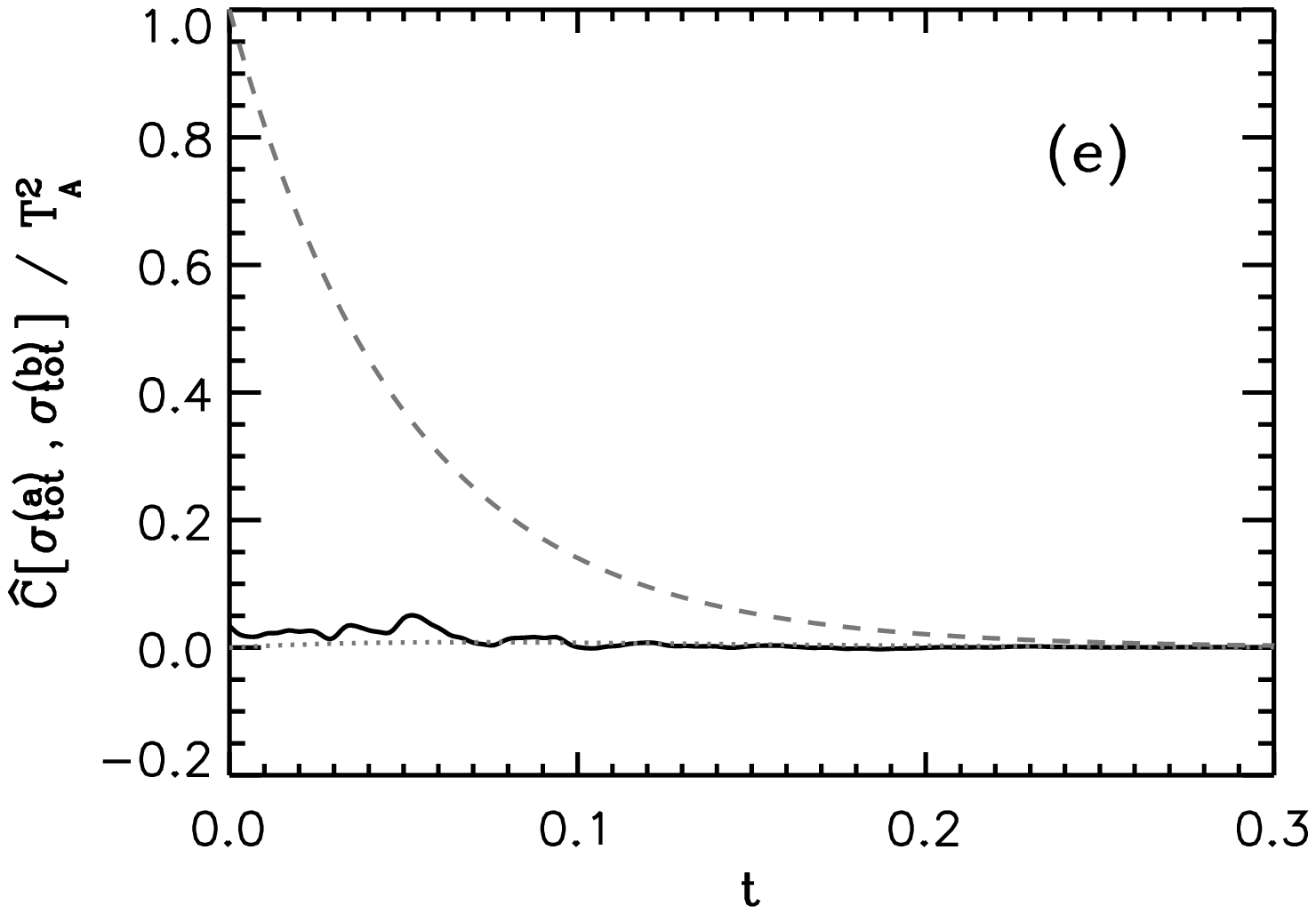}
  \includegraphics[width=0.48\textwidth]{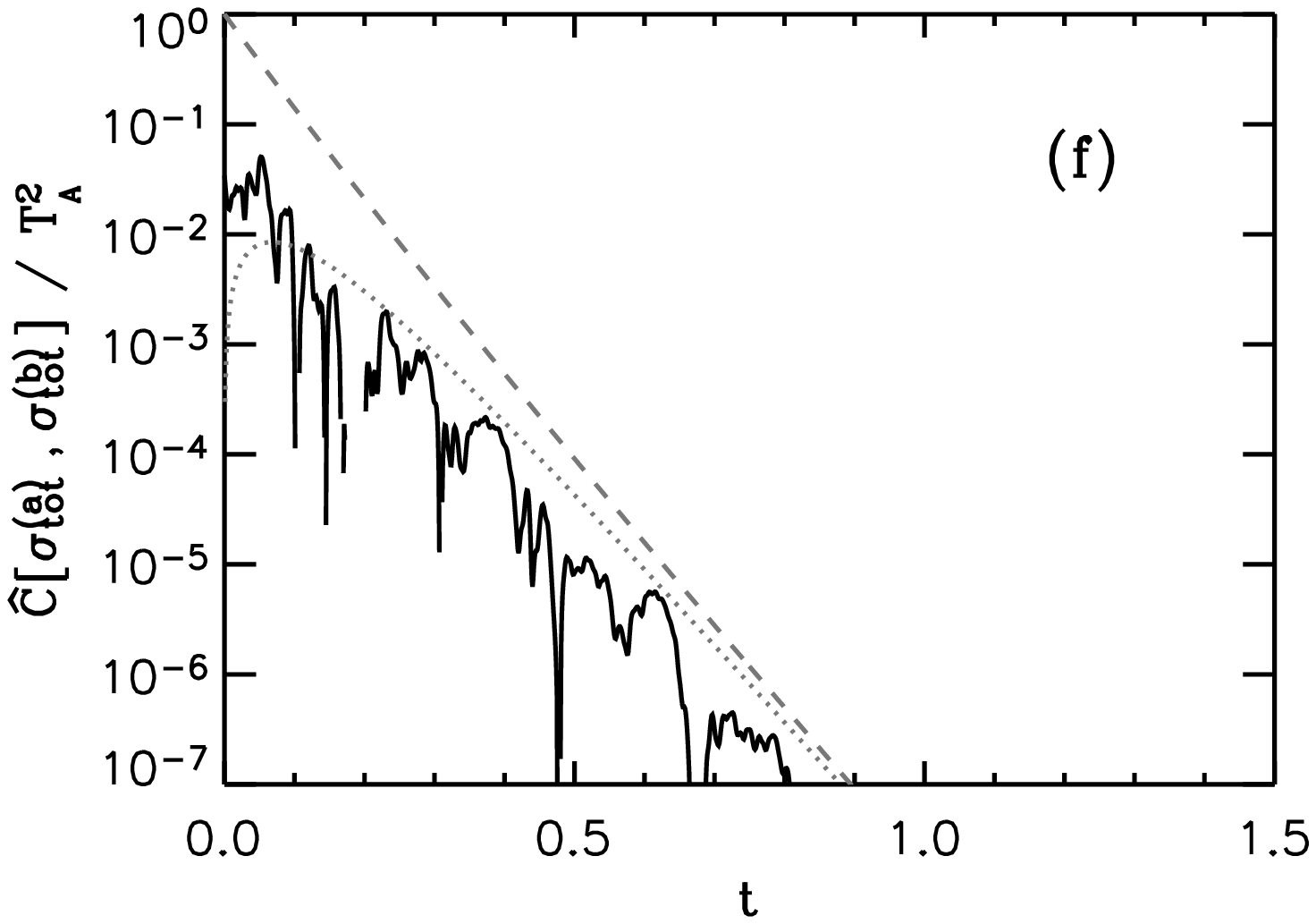} }
 \caption{Cross-correlation function $\hat C[\sigma_{tot}^{(a)},\sigma_{tot}^{(b)}] / T_A^2$ for the fully chaotic billiard, plotted linearly (left column) and logarithmically (right column), where the Fourier transform was taken over three different frequency regimes.
(a), (b): $\nu$=8.5 to 9.5\,GHz, $T_A=0.234$, $T_W=6.79$; 
(c), (d): $\nu$=11 to 12\,GHz, $T_A=0.52$, $T_W=8.26$; 
(e), (f): $\nu$=14 to 15\,GHz, $T_A=0.848$, $T_W=14.92$.
The RBWA is shown for the GOE (dotted line) and for the POE (dashed line).
}
 \label{fig:xcorlog}
\end{figure}

Figure \ref{fig:xcorlog} shows the smoothed results for the fully chaotic
billiard both in linear and logarithmic plots for three different frequency
regimes.
In contrast to the expectation for integrable systems, the cross
correlation is suppressed for small times $t$. The measurement thus
clearly exhibits the correlation hole expected for chaotic systems. In
particular the logarithmic plots demonstrate that the experimental results are
in good agreement with the rescaled Breit-Wigner approximation for the GOE over
several orders of magnitude, and clearly distinguishable from the POE expectations.
For higher frequencies the absorption increases, resulting in a higher value for
$T_W$ and thus a sharper decline of the cross correlation.


 \begin{figure}
 \centerline{ \includegraphics[width=0.48\textwidth]{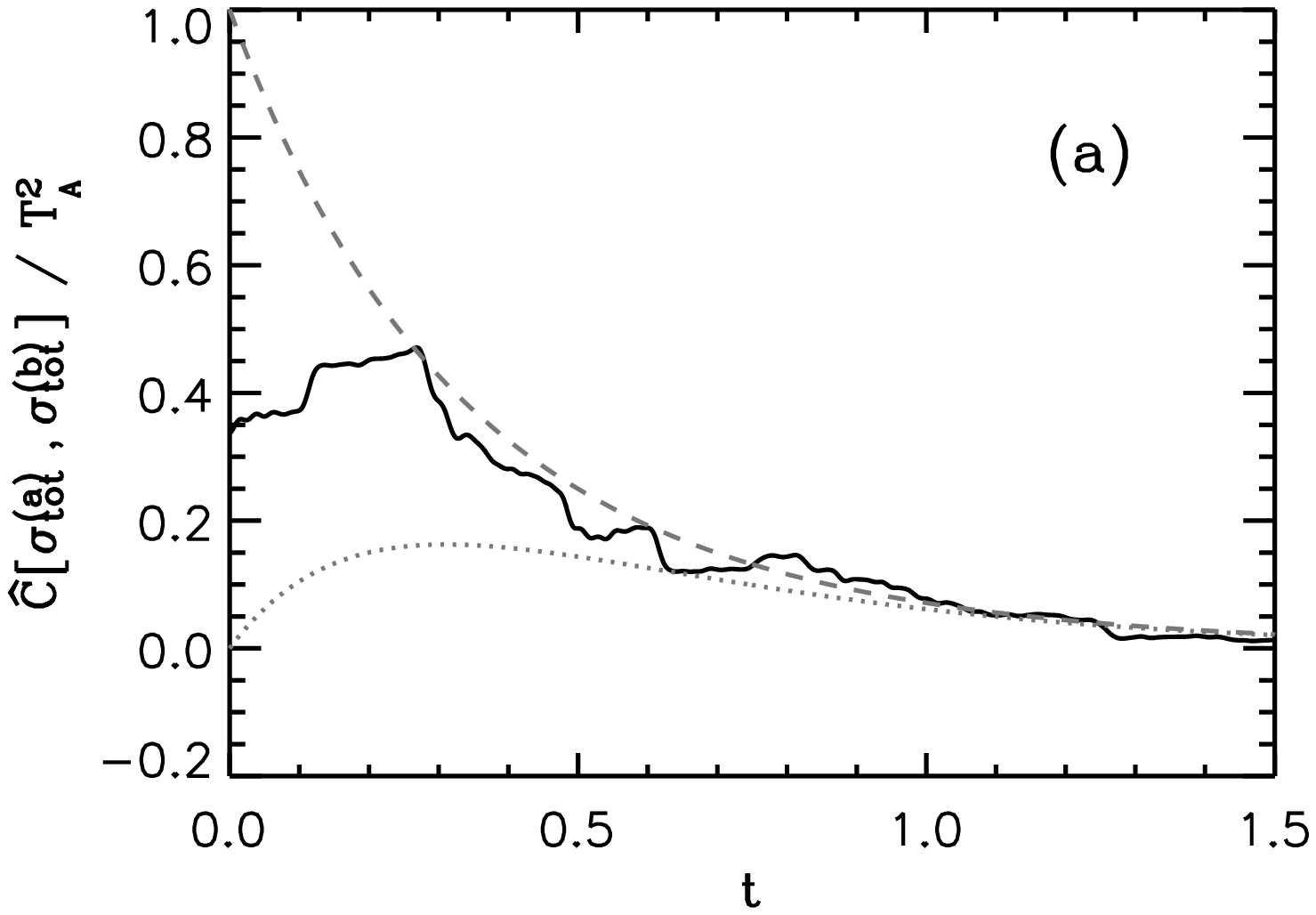}
\includegraphics[width=0.48\textwidth]{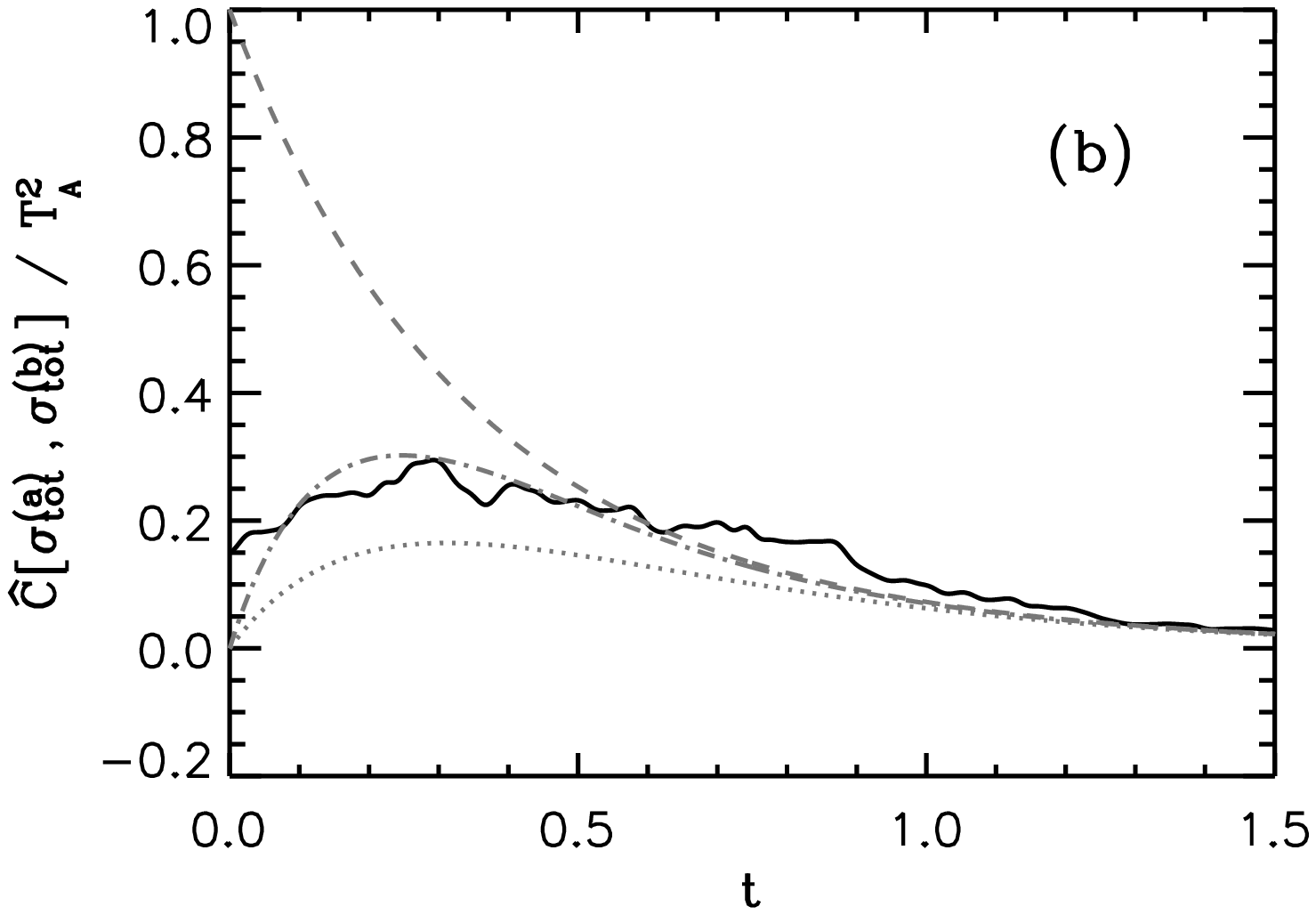}}
 \caption{Cross-correlation function $\hat C[\sigma_{tot}^{(a)},\sigma_{tot}^{(b)}]$ 
for (a) the rectangular and (b) the Robnik billiard with the Fourier transform taken over the frequency range 13 to 14\,GHz.
Dotted and dashed lines are the RBWA results for GOE and POE, respectively.
The dash-dotted line in (b) corresponds to the RBWA for two GOE (see text).
The parameters were $T_A=0.261$, $T_W=1.38$ for the rectangular, and
$T_A=0.254$, $T_W=1.39$ for the Robnik billiard.}
 \label{fig:xcor4}
 \end{figure}

\begin{figure}
 \centerline{ \includegraphics[width=0.48\textwidth]{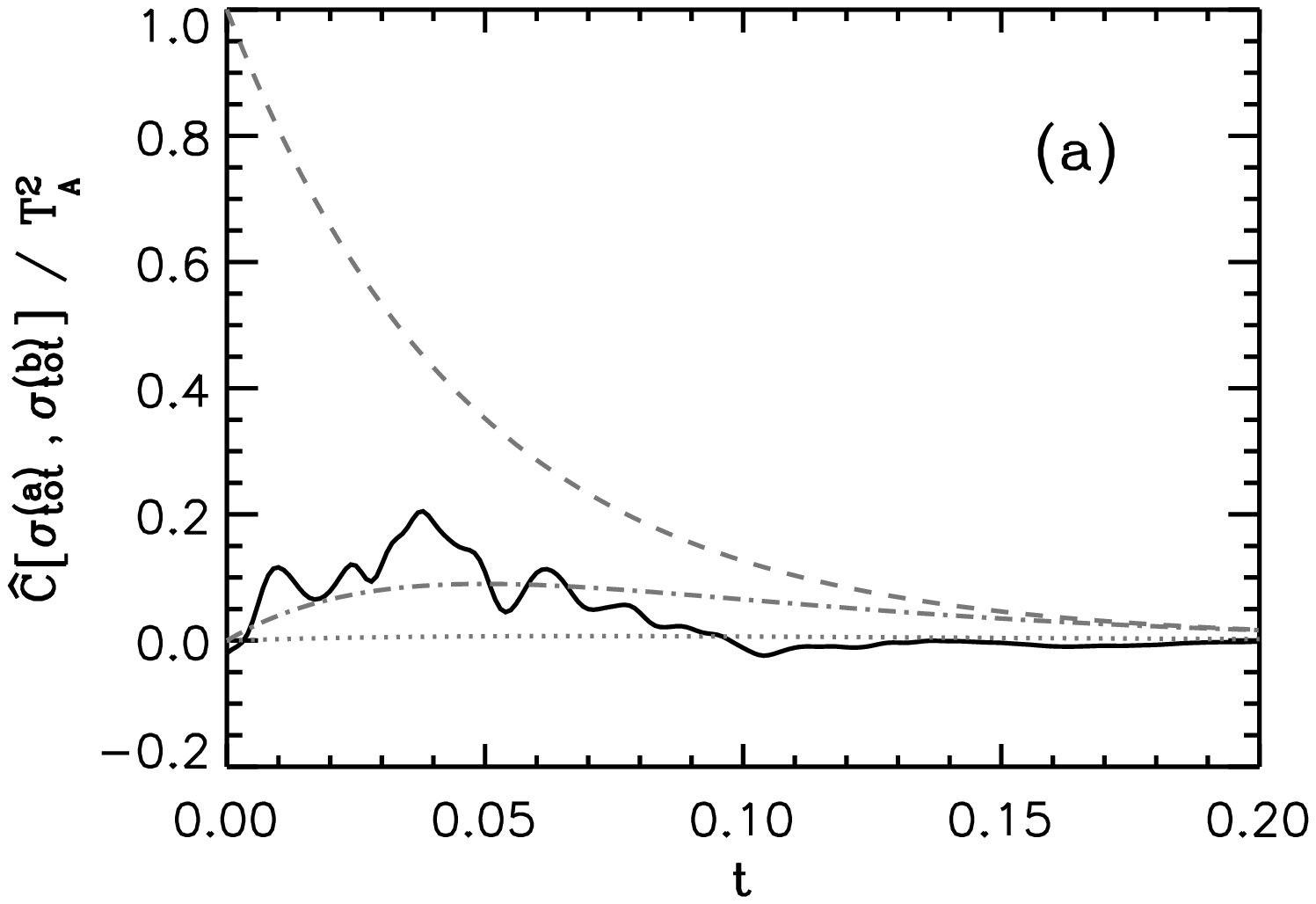}
\includegraphics[width=0.48\textwidth]{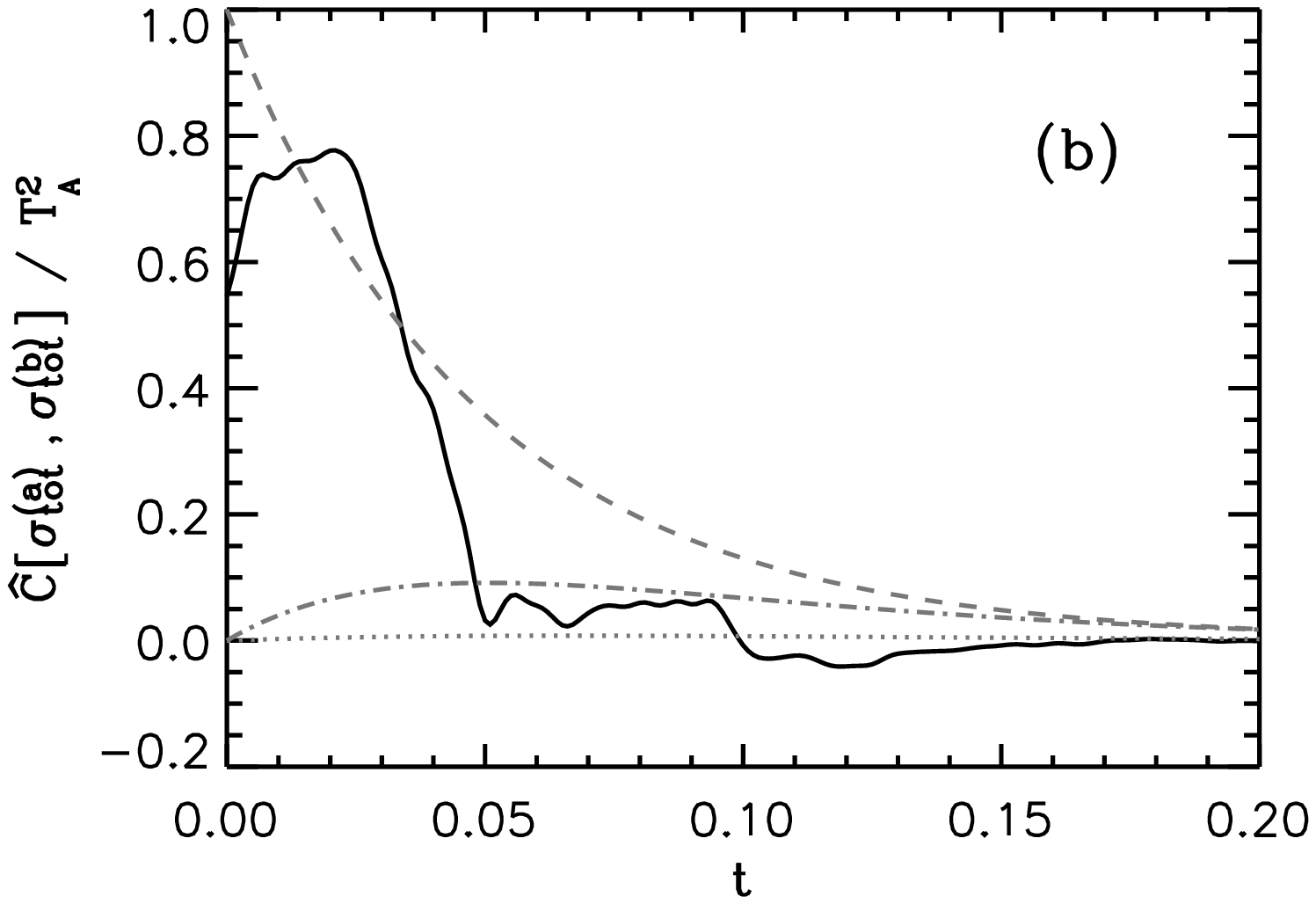}}
 \caption{Cross-correlation function $\hat C[\sigma_{tot}^{(a)},\sigma_{tot}^{(b)}]$ 
for the $C_3$ billiard with (a) asymmetric  and (b) symmetric antenna positions, 
with the Fourier transform taken over the frequency range 14 to 15\,GHz.
Dotted and dashed lines are the RBWA results for GOE and POE, respectively.
The dash-dotted line in (b) corresponds to the RBWA for the $C_3$ billiard (see text). The parameters were $T_A=0.877$, $T_W=15.84$ for (a), and
$T_A=0.860$, $T_W=15.61$ for (b).}
 \label{fig:xcorC3}
 \end{figure}

As an example for an integrable system, we present the cross-correlation function for the
rectangular billiard in figure~\ref{fig:xcor4}(a) together with the rescaled Breit-Wigner
approximation for the GOE and the POE.
The difference to chaotic systems is clearly seen. The discrepancy of our results from 
the POE is not surprising, because the antennas disturb the system, leading to a shift of resonance positions and thus to correlations in the spectrum. 
The intermediate situation is shown in figure~\ref{fig:xcor4}(b) for the Robnik 
billiard with mirror symmetry and non-symmetric antenna positions.
For $\lambda=0.4$ the classical phase space of the Robnik billiard is chaotic \cite{Rob83}, apart from possible tiny stability islands \cite{Dul01}. 
Additionally, the theoretical result for the superposition of two GOE is plotted as dash-dotted line. In this case the two-point form factor $b_2(t)$ for the GOE in equation~(\ref{BreitWigner}) has to be replaced by $b_2(2t)$.

In systems with point symmetries caution is commanded with respect to the antenna positions, 
because the amplitudes at symmetric antenna positions are strongly correlated.
This is illustrated in figure~\ref{fig:xcorC3} showing the cross correlation for the billiard with $C_3$ symmetry both for symmetric and asymmetric antenna positions. In addition to the POE 
and GOE curves, the rescaled Breit-Wigner expectation for the $C_3$ billiard is shown. For the two-point form factor in equation~(\ref{BreitWigner})
the results from reference~\cite{Sch02} are used:
\begin{equation}
b_2^{C3}(t)=\frac{1}{3}\left( b_2^{GOE}(3t)+2 b_2^d\left(\frac{3t}{2}\right) \right) \; ,
\end{equation}
where $b_2^{GOE}(t)$ is the two-point form factor for the singlet GOE spectrum and 
\begin{equation}
b_2^d(t)=-\rme^{-8\pi^2\Delta t^2}+2\rme^{-4\pi^2\Delta t^2}b_2^{GUE}(t)
\end{equation}
is the one for the doublet GUE spectrum. The parameter $\Delta=0.125$ accomodates the splitting of the doublet spectrum due to symmetry breaking.

For asymmetric antenna positions, a good correspondence between experiment and theory is found, but for symmetric positions there are dramatic deviations. In the ideal case one would expect a result which is much closer to the auto-correlation function, but small deviations from symmetry induce uncontrollable variations.

\section{Summary and Outlook}

We have measured the diagonal S-matrix elements of two channels as a function of frequency for 
a variety of different microwave cavities with and without symmetries, and with integrable or chaotic classical dynamics.
They displayed different amounts of resonance overlap and antenna coupling.
Via the optical theorem the diagonal S-matrix elements contain the same information as the total cross sections.

We discuss the experimental results in terms of auto-correlation and cross-correlation functions in the time domain. The wall absorption of the billiards is expressed in terms of unmeasurable channels. We find that an exponential decay of correlations, corresponding to infinitely many channels, describes the experimental results adequately.
Comparison with the rescaled Breit-Wigner approximation for random matrix models shows good agreement with experiment, if we use the wall absorption as a fit parameter.
The cross correlation was expected to display the difference between integrable and chaotic systems more clearly than the auto correlation.
Indeed, our experiments confirm that the correlation hole is more pronounced in the cross-correlation function.

Due to the optical theorem, total cross sections are accessible via the measurement of 
reflection matrix elements for microwave cavities.
This is in contrast to particle scattering experiments, where total cross sections can only be measured in exceptional cases and their measurement in two different entrance channels is even more difficult.
In these experiments only partial cross sections are available. Theory \cite{Gor02} suggests that the correlation hole should be best observable in correlations between cross sections without any coinciding channel indices. The simplest case of this type is the cross-correlation function of different elastic cross sections.

\ack

U Kuhl is thanked for numerous suggestions and helpful discussions. 
Discussions took place at a workshop at the Centro Internacional de 
Ciencias in Cuernavaca, Mexico, in 2001.
TG thanks the EU Human Potential Program contract 
HPRN-CT-2000-00156 for financial support.
THS acknowledges support by CONACyT grant 25192-E, 
and by DGAPA (UNAM) grant IN-109000.
The experiments were supported by the Deutsche Forschungsgemeinschaft.

\section*{References}


\end{document}